\documentclass[superscriptaddress, reprint,nofootinbib, amsmath,amssymb, floatfix]{revtex4-1}

\usepackage[  ]{graphicx} 
\usepackage{dcolumn} 
\usepackage{bm,subfigure} 
 
\newcommand{\rmi}{\mathrm{i}} 
\newcommand{\rmd}{\mathrm{d}} 
 
\newcommand{\rmm}{\mathrm{m}} 
\newcommand{\rmb}{\mathrm{b}} 

\newcommand{\rme}{\mathrm{e}}

\newcommand{\point}{\raise0.7ex\hbox{.}}
\newcommand{\X}{\mathbf{x}}

\newcommand{\rmB}{\mathrm{B}}

\newcommand{\siot}{SiO$_2$ }
\newcommand{\astst}{As$_2$S$_3$ }

\newcommand{\aststnb}{As$_2$S$_3$}

\begin{document}

\preprint{APS/123-QED}

\title{Stimulated Brillouin scattering in   metamaterials}

\author{M. J. A. Smith}
\email{m.smith@physics.usyd.edu.au}
\affiliation{Centre for Ultrahigh bandwidth Devices for Optical Systems (CUDOS) and Institute of Photonics and Optical Science (IPOS), School of Physics, The University of Sydney, NSW 2006, Australia}
 \affiliation{Centre for Ultrahigh bandwidth Devices for Optical Systems (CUDOS), School of Mathematical and Physical Sciences, University of Technology Sydney, NSW 2007, Australia }
\author{B. T. Kuhlmey}
\affiliation{Centre for Ultrahigh bandwidth Devices for Optical Systems (CUDOS) and Institute of Photonics and Optical Science (IPOS), School of Physics, The University of Sydney, NSW 2006, Australia}
\author{C. Martijn de Sterke}
\affiliation{Centre for Ultrahigh bandwidth Devices for Optical Systems (CUDOS) and Institute of Photonics and Optical Science (IPOS), School of Physics, The University of Sydney, NSW 2006, Australia}

\author{C. Wolff}
\author{M. Lapine}
\author{C. G. Poulton}
 \affiliation{Centre for Ultrahigh bandwidth Devices for Optical Systems (CUDOS), School of Mathematical and Physical Sciences, University of Technology Sydney, NSW 2007, Australia }

\date{\today} 

\begin{abstract} \noindent
We compute the SBS gain for a  metamaterial comprising a cubic lattice of dielectric spheres suspended in a background dielectric material. Theoretical methods are presented to calculate the  optical, acoustic, and opto-acoustic parameters that describe the SBS properties of the  material at long   wavelengths.  Using the electromagnetic and strain energy densities we accurately characterise the optical and acoustic properties of the metamaterial. From a combination of  energy density methods and perturbation theory, we   recover the appropriate terms of the photoelastic tensor for the metamaterial. We demonstrate that   electrostriction is not necessarily the dominant mechanism in the enhancement and suppression of the  SBS gain coefficient in a metamaterial, and that other parameters, such as the Brillouin linewidth, can dominate instead. Examples are presented that exhibit an order of magnitude enhancement in the SBS gain as well as perfect suppression.

\end{abstract}

\maketitle

\section{Introduction}
 Stimulated Brillouin Scattering (SBS) is a nonlinear opto-acoustic   process by which an incident optical pump field       generates   an acoustic wave inside a  dielectric material. The longitudinal acoustic wave compresses the medium periodically to form a diffraction grating, which scatters the incident field; due to the acoustic wave propagation, the scattered field is Doppler shifted to lower frequencies. This amplifies the beating between the Stokes field and the pump field, creating  a self-reinforcing effect. Although SBS was theoretically predicted by Brillouin in 1922 \cite{brillouin1922diffusion}, it was only after the advent of the laser in 1960 \cite{maiman1960stimulated} that it could be practically realised, with the first experimental paper in SBS following soon after in 1964 \cite{chiao1964stimulated}. More recently,   research  focus has been directed towards the design of practical, small-scale devices that exploit SBS, such as notch-filters,  Brillouin lasers, and   microwave sources \cite{pant2014chip,morrison2014tunable,kabakova2013narrow,eggleton2013inducing}. For such device applications, there is considerable interest in  materials which exhibit strong SBS, as this allows  improved power scaling and subsequently, smaller devices. At the same time,  SBS    is also regarded as a   nuisance in the optical fibre community \cite{peral1999degradation} where  SBS     acts as a power limit for narrow linewidth signals sent through fibres,  and hinders the observation of other nonlinear effects such as four-wave mixing \cite{powers2011fundamentals} at power levels above the SBS threshold. Although there are established experimental procedures for overcoming SBS, such as frequency dithering \cite{agrawal2007nonlinear},  there is considerable interest in designer   materials  which exhibit intrinsically high, or in certain circumstances, intrinsically low  SBS gain.

Here we   examine the effects of subwavelength structuring on the intrinsic (bulk) SBS gain spectrum of a material. This work provides an in-depth analysis following our initial observations  \cite{smith2016control}  in which it was shown that  considerable enhancement or complete suppression of the SBS gain can be achieved in a simple cubic lattice of spheres in a background material.  In order to characterise the SBS properties of a metamaterial it is necessary to determine    a combination of optical, acoustic, and opto-acoustic parameters that feature in the SBS gain spectrum expression \cite{powers2011fundamentals}; here we give a rigorous formalism whereby these parameters can be computed. To obtain the   optical properties (the effective refractive index) we use a long-wavelength homogenisation procedure based on the  optical energy density \cite{bergman1978dielectric}. For independent validation, we also outline an equivalent procedure which uses the slope of the lowest band surface near the centre of the first Brillouin zone \cite{movchan2002asymptotic}.  To obtain the  acoustic properties of the metamaterial we  present a long-wavelength procedure, this time using   the acoustic energy density, to determine the   stiffness tensor and subsequently the longitudinal acoustic velocity for the metamaterial. The remaining acoustic parameters including the frequency shift and  line width are obtained by incorporating acoustic damping as a perturbation to the  acoustic wave equation. In this way we incorporate the   effects of acoustic loss, which is critical to the SBS process, to leading order. The procedure for the opto-acoustic parameter (an element of the fourth-rank photoelastic tensor $p_{ijkl}$ \cite{newnham2004properties}), whilst conceptually simple, is the most technically difficult to obtain and relies on a combination of optical homogenisation and  acoustic  boundary perturbations.

The solution procedure used makes very few assumptions on the mechanical and optical response of the material, assuming primarly that the constituents are linear elastic dielectric materials in a three-dimensional Bravais lattice configuration.  We demonstrate the method for metamaterials comprising a three-dimensional cubic lattice of dielectric spheres embedded in a dielectric background, including simple cubic (sc) and face-centred cubic (fcc)  configurations, with a single element per unit cell. These metamaterial designs are chosen not only due to their geometrical elegance, but because they possess closed form expressions in the dilute asymptotic limit, which can be used for numerical validation with suitable approximations.   The theoretical framework outlined in the present work is   relatively general   and can be used to calculate the SBS properties of a wide class of metamaterials.

The outline of this paper is as follows: in Section \ref{sbs:iso} we present a brief overview of SBS in isotropic materials, outlining the key material and wave parameters necessary to determine the SBS gain. The following sections then consider techniques to determine these parameters for a metamaterial: In Section \ref{sec:effn} we outline methods to determine the permittivity tensor, in Section \ref{sec:mechparams} we present techniques to determine the purely acoustic terms, and in Section \ref{sec:photoel} we investigate the photoelastic properties. This is followed in Section \ref{sec:numer} by a selection of numerical examples and a discussion in Section \ref{sec:concl}. Appendix \ref{sec:effperm} discusses  the formal definition of the effective permittivity tensor and Appendix \ref{sec:numimp} is an extended comment on the numerical implementation for Section \ref{sec:photoel}. 
 
  \section{SBS in isotropic materials} \label{sbs:iso}
 We consider the theoretical framework for an electrostriction-driven backward  SBS process in an optically and acoustically isotropic bulk material that exhibits negligible optical losses. Acoustic  losses are assumed to be well-modelled  through the inclusion of dynamic viscosity effects in the model (linear friction in continuum mechanics), and this is taken to be the dominant loss mechanism  for acoustic problems \cite{auld1973acoustic}. We also neglect a magnetic response in the metamaterial   and   additional nonlinear optical effects, such as four-wave mixing. In this setting,  we define the total electric field as   $  \mathbf{E} = (0,0,E_z)$ with
\begin{equation}
\label{eq:elecfield}
E_z =   \frac{1}{2}\left( A_1 \rme^{\rmi k_1 z - \rmi \omega_1 t} +A_2 \rme^{-\rmi k_2 z - \rmi \omega_2 t}  \right) + \mathrm{c.c.},
\end{equation}
 where $A_{1,2}$ are the   amplitudes, $k_{1,2} = n \omega_{1,2} / c_0$  the wave numbers,   $\omega_{1,2}$   the angular frequencies  of the incident pump field and scattered (Stokes) field, respectively, and $c_0$ is the speed of light in vacuum. The system that describes the intensities of the incident pump and Stokes field takes the form \cite{agrawal2007nonlinear,peral1999degradation,kobyakov2010stimulated}  
 \begin{subequations}
\begin{align}
 \partial_z I_1  &= -g_\mathrm{P} I_1 I_2,\\
 \partial_z I_2 &= g_\mathrm{P} I_1 I_2,
\end{align}
 \end{subequations}
 where  $I_j =   \varepsilon_0 n^\mathrm{ }  c_0 A_j A_j^\ast /2$,  $\varepsilon_0$  is the vacuum permittivity, and  $n$ is the refractive index.   The SBS power gain spectrum is given by   \cite{kobyakov2010stimulated,powers2011fundamentals}
\begin{equation}
\label{eq:gain}
g_\mathrm{P}(\Omega) = \frac{ 4 \pi^2  n^7 p_\mathrm{xxyy  }^2   }{   c_0 \lambda_1^2 \rho   \nu_\mathrm{A} \Gamma^\mathrm{ }_\rmB} \left( \frac{( \Gamma^\mathrm{ }_\rmB/2)^2}{(\Omega_\rmB^\mathrm{ } - \Omega)^2 + ( \Gamma^\mathrm{ }_\rmB/2)^2}\right),
\end{equation}
where   $p_\mathrm{xxyy} $ is the relevant component of the   elasto-optic (photoelastic, or, Pockels)   tensor $p_{ijkl}$ for SBS in bulk media, $\Gamma_\rmB$  denotes the Brillouin line width  with respect to angular frequency, $\lambda_1$  is the wavelength for frequency $\omega_1$  in vacuum, $\rho$ is  the mass density,  $\nu_\mathrm{A}$ is the   acoustic wave velocity, and  $\Omega/(2\pi)$ is the     acoustic wave frequency.

 At the Brillouin resonance,  a backwards travelling  longitudinal acoustic wave is excited with angular frequency $\Omega_\rmB$, phase velocity $\nu_\mathrm{A}$ and   acoustic wave vector $\mathbf{q}_\rmB$. From  conservation of momentum, and noting that $c_0 \gg \nu_\mathrm{A}$, we have that
  \begin{subequations}
\begin{equation}
\label{eq:acoustblochvec}
|\mathbf{q}_\rmB| = |\mathbf{k}_1| + |\mathbf{k_2}| \approx 2 |\mathbf{k}_1|, \\
\end{equation}
and subsequently from the dispersion relations for  the optical and acoustic problems it follows that
 \begin{equation}
\label{eq:energyrel}
\Omega_\rmB   \approx \frac{2 \omega_1 n  \nu_\mathrm{A}}{c_0},
\end{equation}
 \end{subequations}
To summarise, knowledge of the material density $\rho$, the refractive index $n$, the acoustic parameters $\Omega_\rmB,\Gamma_\rmB$,  and  $\nu_\mathrm{A}$,  and the photoelastic tensor element $p_\mathrm{xxyy}$,   allows us to determine the bulk SBS power gain at a specified optical wavelength $\lambda_1$. For reference, a table of   bulk material parameters for a range of common materials can be found in \cite{smith2016control}. To the best of the authors knowledge, there is no general theoretical treatment  of bulk SBS in   anisotropic media, however \cite{wolff2014germanium} has examined SBS in cubic media.

\section{  Optical properties} \label{sec:effn}
There are an extensive number of  procedures which may be used to determine the   optical properties of a composite material  at long wavelengths \cite{milton2002theory}. We outline   two methods for determining the effective refractive index; the first having the advantage of being conceptually simple and numerically efficient, but is an indirect method and is only valid for dielectric metamaterials possessing cubic symmetry, and the second being a much more general tool which homogenises the material as opposed to describing wave propagation.

The first and most commonly used approach,   is to consider   the optical wave equation  
\begin{subequations}
\begin{equation}
\label{eq:maxwell}
\nabla \times  \boldsymbol{\varepsilon}_\mathrm{r}^{-1} \nabla \times \mathbf{H}  = \left( \frac{\omega}{c_0}\right)^2 \mathbf{H},
\end{equation}
with Bloch--Floquet boundary conditions   on the edges of the fundamental cell
\begin{equation}
\label{eq:bfconds}
\mathbf{H}(\mathbf{x} + \mathbf{R}_p)\big|_{\partial W} = \mathbf{H}(\mathbf{x}) \, \rme^{\rmi \mathbf{k} \cdot \mathbf{R}_p}\big|_{\partial W}\,,
\end{equation}
and where the     tangential components of $\mathbf{E}$ and $\mathbf{H}$   are continuous across all dielectric interfaces $\partial U_j$:
\begin{equation}
\label{eq:bcboundary}
\mathbf{n} \times   \ulcorner  \mathbf{H} \lrcorner \big|_{\partial U_j} = 0, \quad \mathbf{n} \times   \ulcorner  \mathbf{E} \lrcorner \big|_{\partial U_j} = 0.
\end{equation}
\end{subequations}
Here $\boldsymbol{\varepsilon}_\mathrm{r}$ is the matrix representation of the relative permittivity tensor $\varepsilon_{ij}$, $\mathbf{H}$ denotes the  magnetic field   of the Bloch mode, $\mathbf{n}$ is the normal vector to the boundary $\partial U_j$ of the $j$th inclusion, $\ulcorner \lrcorner$ denotes the jump discontinuity in the field across the interface,  $\mathbf{x} = (x,y,z)$ is defined with respect to a conventional Cartesian basis, $\mathbf{R}_p$ is the real lattice vector, and   ${\partial W}$ denotes the edges of the entire fundamental cell. For a cubic lattice  $\mathbf{R}_p = d (i,j,k)$  for $(i,j,k) \in \mathbb{Z}_3$, and $d$ is the period.

The optical boundary value problem described by \eqref{eq:maxwell}-\eqref{eq:bcboundary}  is then numerically solved for    a Bloch vector $\mathbf{k} $ sufficiently close to the $\Gamma$ point, i.e. $|\mathbf{k}| \ll \pi/d$. Using the  lowest angular eigenfrequency $ {\omega}( {\mathbf{k}})$ the effective refractive index is obtained via
\begin{equation}
\label{eq:neffexpr}
n^\mathrm{eff} = \frac{c_0 | {\mathbf{k}}|}{ {\omega}}.
\end{equation}
This   procedure only applies to cubic metamaterials, and is an indirect method for characterising the optical properties of the medium. This is because  the refractive index describes wave propagation whereas the permittivity is an intrinsic material property.   Such a shortcoming can  easily be overcome by considering an alternative procedure;    a modification of the method outlined in  \cite{bergman1978dielectric} which we term an {\it energy density method}.  We   introduce this   procedure by   considering the approach for a metamaterial of cubic symmetry. As before, it  requires that we solve \eqref{eq:maxwell}-\eqref{eq:bcboundary}  for a  wave vector $  {\mathbf{k}}  $ close to $\Gamma$, except we  now   compute the volume-averaged electromagnetic energy density
 \begin{subequations}
\begin{equation}
\label{eq:Uavg}
  \mathcal{E}_\mathrm{avg} = \frac{1}{2} \frac{1}{V_\mathrm{WSC}} \varepsilon_0 \int_{W} \varepsilon_\mathrm{r} |\mathbf{E}|^2 \, \rmd W,
\end{equation}
where  $V_\mathrm{WSC}$ is the volume of the Wigner-Seitz cell \cite{kittel2005introduction}. This quantity is then equated to an   effective  energy density       ansatz
\begin{equation}
\label{eq:Ueff}
     \mathcal{E}_\mathrm{eff}   =  \frac{1}{2} \frac{1}{\left(V_\mathrm{WSC}\right)^2}  \varepsilon_0  \varepsilon_\mathrm{r}^\mathrm{eff}    \left|\int_{W}  \mathbf{E} \, \rmd W \right|^2,
   \end{equation}
to obtain the scalar effective permittivity  
\begin{equation}
\varepsilon_\mathrm{r}^\mathrm{eff} = V_\mathrm{WSC} \frac{ \int_{W} \varepsilon_\mathrm{r} |\mathbf{E}|^2 \, \rmd W}{    \left|\int_{W}  \mathbf{E} \, \rmd W \right|^2}.
\label{eq:unifeffperm}
\end{equation}
 \end{subequations}
 Since our cubic metamaterial comprises dielectric media and is examined in the long wavelength limit, we take $\mu_\mathrm{r}^\mathrm{eff} =1$ throughout and so    $ n^\mathrm{eff} =( \varepsilon_\mathrm{r}^\mathrm{eff})^{1/2} $ follows immediately.  We remark that there is no difference in the numerical result obtained using either  \eqref{eq:neffexpr} or \eqref{eq:unifeffperm} . The  equivalence of the two effective medium methods outlined in this section can be found in Appendix \ref{sec:effperm}, where it is shown that the equivalence holds provided  $\mathbf{k}$ is parallel to a principal axis vector.  The primary motivation for using \eqref{eq:unifeffperm} can be found in later sections, when we require the  effective permittivity tensor for the metamaterial under an applied  strain.  When the metamaterial is strained the symmetry class changes, and subsequently a generalisation of \eqref{eq:unifeffperm} is required. By evaluating the permittivity tensor directly, we avoid the added step of reverse engineering the   effective permittivity tensor from multiple scalar refractive index values.

The generalisation of \eqref{eq:unifeffperm} for other lattice configurations takes the form
\begin{multline}
\sum_{j \in \left\{x,y,z \right\}} \widetilde{\varepsilon}_{jj}^\mathrm{\;eff}  \bigg\vert\int_{W}  E_j \, \rmd W \bigg\vert^2  
 \\
=V_\mathrm{WSC }    \int_{W}   \sum_{i,j \in \left\{x,y,z \right\}} \widetilde\varepsilon_{ij}  E_i E_j^\ast \, \rmd W,
\label{eq:epseffsystem}
\end{multline}
where $ \widetilde{\varepsilon}_{jj}$ denotes  principal dielectric constants \cite{born1964principles}. These constants  are defined with respect to the principal dielectric axes  of the metamaterial.   In order to obtain an invertible linear system for the three unknown $ \widetilde{\varepsilon}_{jj}$,  we consider three different Bloch vectors     near the $\Gamma$ point that are parallel to the   principal dielectric axes of the metamaterial.  Once the principal permittivities are obtained, the   effective permittivity tensor $\varepsilon_{ij}^\mathrm{eff}$ is obtained via a change of basis operation, as the effective permittivity   is defined with respect to  the Cartesian coordinate axes. For cubic and tetragonal metamaterials, for example,  it can be shown that $ \widetilde{\varepsilon}_{ij}    = \varepsilon_{ij} $. Note that the principal dielectric axes of monoclinic and triclinic metamaterials cannot be predicted {\it a priori} and in such instances this procedure cannot be used.

\section{  mechanical properties} \label{sec:mechparams}
 Next we consider the   mechanical parameters of our metamaterial, and proceed with the simplest of these; determining the   material density, which is given by simple volume averaging. Explicitly, given the   filling fraction $f$ of the inclusion material in the  background material, the   density       is given by
\begin{equation}
\rho^\mathrm{eff} = \rho_\rmi f + \rho_\rmm (1-f),
\end{equation}
where $\rho_{\rmi,\rmb}$ denote the mass densities of the inclusion and background material, respectively.

We now compute the acoustic phase velocity at long wavelengths, where in the absence of acoustic loss, the governing equation  is \cite{auld1973acoustic}
 \begin{subequations}
\begin{equation}
\label{eq:acwaveeq}
\rho \Omega^2 \mathbf{u} + \nabla \cdot (\mathbf{C}:   \mathbf{s}) = 0,
\end{equation}
 where $\mathbf{u}$ denotes the mechanical displacement from equilibrium, $\mathbf{C}$ is the fourth-rank stiffness tensor,   $\mathbf{s}$ is the strain tensor,  (defined   as the symmetric gradient of the displacement: $s_{ij} =  (\partial_i u_j + \partial_j u_i)/2$), the symbol $:$ denotes the inner product between a fourth- and second-rank tensor,  and we    consider  time harmonic solutions of the form $\mathrm{exp}(-\rmi \Omega t)$. \eqref{eq:acwaveeq} is solved     with  the  Bloch  condition
\begin{equation}
\label{eq:bfacoustic}
\mathbf{u}(\mathbf{x} + \mathbf{R}_p)\big|_{\partial W} =  \mathbf{u}(\mathbf{x}) \rme^{\rmi \mathbf{q} \cdot \mathbf{R}_p }\big|_{\partial W},  
\end{equation}
 which is defined  with respect to the acoustic Bloch vector $\mathbf{q}$, and  assuming continuity of the mechanical displacement field and vanishing normal stress  across all mechanical interfaces 
\begin{equation}
\label{eq:ctbcacoustic}
\ulcorner \mathbf{u} \lrcorner \big|_{\partial U_j} = 0, \quad \ulcorner \boldsymbol{\sigma} \lrcorner \big|_{\partial U_j} \cdot \mathbf{n} = 0,
\end{equation}
\end{subequations}
where $\boldsymbol{\sigma}$ denotes the stress tensor. It is from the boundary-value problem described in \eqref{eq:acwaveeq}--\eqref{eq:ctbcacoustic} above that we determine the   phase velocity for the longitudinal acoustic wave excited by SBS in a metamaterial. A conventional approach, which is valid when the material is structured to form a cubic crystal, is to simply consider an acoustic wave vector at the SBS resonance  $\widetilde{\mathbf{q}} = (0,0,4 \pi n^\mathrm{eff} / \lambda_1)$  and  use the     angular frequency $\widetilde{\Omega}$ corresponding  to the approximately longitudinal mode, to obtain
  \begin{equation}
  \nu_\mathrm{A}^\mathrm{eff} = \frac{\widetilde{\Omega}}{|\widetilde{\mathbf{q}}|}.
  \end{equation}
This method (which implicitly assumes that $\widetilde{\mathbf{q}}$  is also sufficiently close to the $\Gamma$ point) is only applicable to  highly symmetric problems, and accordingly, a more general approach is required.

 The procedure we consider here is   analogous to that used for determining the effective permittivity   in Section \ref{sec:effn}, except  we now consider the elastodynamic energy density to determine the effective fourth-rank stiffness  tensor for the metamaterial. For a given acoustic Bloch vector $\widetilde{\mathbf{q}}$ we compute the volume-averaged strain energy density  
 \begin{subequations}
\begin{equation}
\label{eq:acenden}
\mathcal{E}_\mathrm{avg}^\mathrm{a} =
 \frac{1}{V_\mathrm{WSC}} \sum_{i,j,k,l \in \left\{ \mathrm{x},\mathrm{y},\mathrm{z} \right\}} \int_W \left\{ s_{ij} C_{ijkl} s_{kl}^\ast \right\} \, \rmd W
\end{equation}
and equate this to  an effective strain energy density ansatz
\begin{equation}
\mathcal{E}_\mathrm{eff}^\mathrm{a} =
 \frac{1}{V_\mathrm{WSC}^2} \sum_{i,j,k,l \in \left\{ \mathrm{x},\mathrm{y},\mathrm{z} \right\}} \left( \int_W s_{ij} \, \rmd W \right)  C_{ijkl}^\mathrm{eff} \left( \int_W s_{kl} \, \rmd W \right)^\ast
\end{equation}
to obtain the system
\begin{multline}
\label{eq:effepsilonrfull}
 \sum_{i,j,k,l \in \left\{ \mathrm{x},\mathrm{y},\mathrm{z} \right\}}  C_{ijkl}^\mathrm{eff} \left( \int_W s_{ij} \, \rmd W \right)  \left( \int_W s_{kl} \, \rmd W \right)^\ast \\
=  V_\mathrm{WSC}  \sum_{i,j,k,l \in \left\{ \mathrm{x},\mathrm{y},\mathrm{z} \right\}} \int_W \left\{ s_{ij} C_{ijkl} s_{kl}^\ast\right\} \, \rmd W.
\end{multline}
 \end{subequations}
Assuming symmetric stress and strain tensors,  reversible deformations, and a metamaterial with cubic symmetry, the total number of unknown   coefficients is reduced from 81 to 3   (these are $C_\mathrm{xxxx}^\mathrm{eff}$, $C_\mathrm{xxyy}^\mathrm{eff}$, and $C_\mathrm{yzyz}^\mathrm{eff}$) making the calculation in \eqref{eq:effepsilonrfull} above   tractable.

Having determined the   $C_{ijkl}^\mathrm{eff}$ coefficients, expressions for the acoustic phase velocities (quasi-longitudinal $\nu_\mathrm{A}^\mathrm{L}$, quasi-shear $\nu_\mathrm{A}^\mathrm{QS}$ and pure shear $\nu_\mathrm{A}^\mathrm{S}$) follow from the acoustic dispersion equation of an isotropic, cubic material. For  any cubic  material with an acoustic wave vector  oriented along a crystal axis we have \cite{auld1973acoustic}
\begin{equation}
\nu_\mathrm{A}^\mathrm{L} = \sqrt{\frac{C_\mathrm{xxxx}}{\rho}}, \quad \mbox{and} \quad \nu_\mathrm{A}^\mathrm{S} = \nu_\mathrm{A}^\mathrm{QS} =  \sqrt{\frac{C_\mathrm{yzyz}}{\rho}}.
\end{equation}
where the longitudinal wave speed is the velocity term present in the SBS gain coefficient.

We now evaluate the two remaining mechanical parameters necessary to evaluate the   SBS gain coefficient in \eqref{eq:gain} for a metamaterial: the Brillouin frequency shift $\Omega_\rmB$ and Brillouin line width $\Gamma_\rmB$. These parameters are obtained by incorporating the effects of mechanical loss into \eqref{eq:acwaveeq} above. In general terms, the inclusion of loss alters the eigenvalues of the acoustic wave equation to make them complex-valued, with the frequency shift and line width given by the real and imaginary components of the appropriate eigenvalue (see \eqref{eq:OmegaPrime} below). For an SBS process, the acoustic mode is purely longitudinal and so we consider   the acoustic frequency which lies on  the first longitudinal band surface at the resonant acoustic wave vector $\widetilde{\mathbf{q}}$. We begin by assuming  that     mechanical losses can be  modelled, at least  to leading order,  by  including   phonon  viscosity effects in our model.  Subsequently,  $\mathbf{C}$ is replaced by   $\mathbf{C} + \partial_t \boldsymbol{\eta}$  in the   acoustic wave equation \eqref{eq:acwaveeq}  where    $\boldsymbol{\eta}$ denotes the fourth-rank dynamic phonon viscosity tensor.  Provided $|\partial_t {\eta}_{ijkl}| \ll | {C}_{ijkl}|$   for all indices, we   treat the effect of phonon viscosity as a perturbation to the acoustic frequencies of the lossless mechanical problem. If we take  the dot product of \eqref{eq:acwaveeq} with $\mathbf{u}^\ast$ and integrate over the unit cell, the eigenvalues $ {\Omega}^2$   take the form $( {\Omega}^\prime)^2$ where
\begin{equation}
\left(  {\Omega}^\prime\right)^2 =   {\Omega}^2 - \rmi  {\Omega} \frac{\int_W \, \nabla \cdot \left( \boldsymbol{\eta} : \nabla \mathbf{u} \right) \cdot \mathbf{u}^\ast \rmd W}{\int_W \rho \mathbf{u} \cdot \mathbf{u}^\ast \, \rmd W}.
\end{equation}
The Brillouin   frequency shift and the acoustic damping   are obtained from the   longitudinal acoustic frequency by simply evaluating a square root to obtain 
\begin{equation}
\label{eq:OmegaPrime}
\widetilde{\Omega}^\prime = \Omega_\mathrm{B}  - \rmi \frac{\Gamma_\mathrm{B} }{2}.
\end{equation}
  Note that we have used the convention of examining complex frequency and real Bloch vector, which follows from our     representation of the SBS gain coefficient in \eqref{eq:gain} (for a formalism in terms of the incident wave vector, see \cite{wolff2014stimulated}).

\section{  Photoelastic properties} \label{sec:photoel}
\begin{figure}[t]
\begin{center}
  \includegraphics[width=0.48\textwidth]{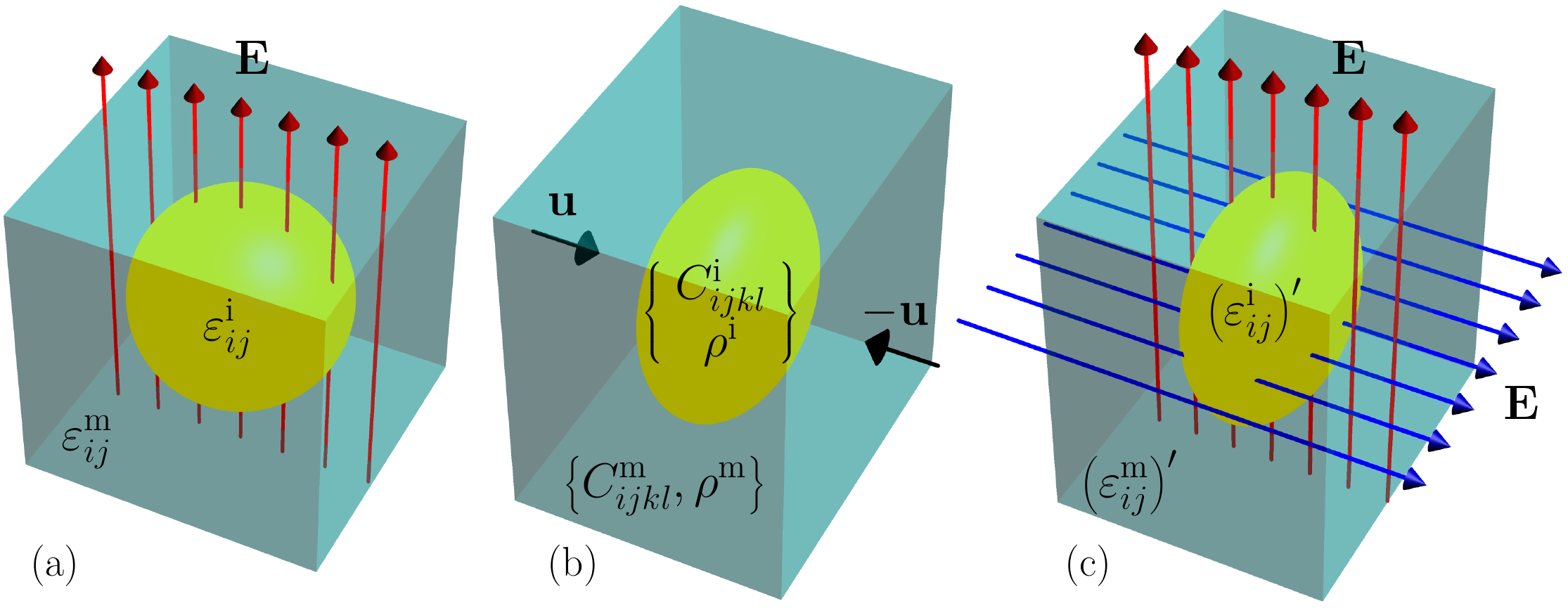}
\caption{Schematic   for computing $p_\mathrm{xxyy}^\mathrm{eff}$: (a) determine   effective index for   unstrained configuration, (b) compute   deformed geometry for   macroscopic strain field $s_\mathrm{yy}^\mathrm{eff}$, (c) determine   effective index for   strained configuration, which requires two wave vector directions.}
\label{fig:3figmerge}
\end{center}
\end{figure}

In this section we determine the photoelastic tensor element $p_\mathrm{xxyy}^\mathrm{eff}$ present in the gain spectrum expression \eqref{eq:gain}. The photoelastic tensor of a material is   implicitly defined via 
\begin{equation}
\label{eq:photoelgen}
\delta \boldsymbol{\varepsilon}_\mathrm{r}^{-1} = \mathbf{p} : \mathbf{s},
\end{equation}
where  $\delta \boldsymbol{\varepsilon}_\mathrm{r}^{-1} $ denotes the change in the inverse relative permittivity tensor and $\mathbf{p}$ denotes the fourth-rank photoelastic tensor. The procedure we use to determine $p_\mathrm{xxyy}^\mathrm{eff}$ for a cubic metamaterial is presented schematically in Figure \ref{fig:3figmerge} and we outline this in   detail.

First, we compute the effective inverse permittivity tensor for the   unstrained unit cell   $(\varepsilon_{ij}^{-1})^\mathrm{unstr,eff}$ (Figure \ref{fig:3figmerge}(a)).

Second, we  compress the unit cell  by imposing   displacements on the cell boundary to obtain  the strained unit cell and the internal strain field   (Figure \ref{fig:3figmerge}(b)). We solve   the acoustic wave equation \eqref{eq:acwaveeq} in the static limit with the boundary conditions
\begin{equation}
\label{eq:syydispcond}
\mathbf{u}\big|_{ \pm \partial W_y} =   \left( 0, - D \mathrm{y},0\right)^\mathrm{T}, \quad  \mathbf{u} \cdot \mathbf{n} \big|_{\partial W \backslash \pm \partial W_y}= 0,
\end{equation}
where $\pm \partial W_y$ denote the faces of the cubic cell with normal vectors $\mathbf{n} = (0, \pm 1,0)^\mathrm{T}$ and $D \ll 1$ is the positive-valued amplitude of the imposed displacement.  These  boundary conditions give  a   strain across the unit cell of  $s_\mathrm{yy} = -D$.  Under this compression, the inclusion geometries inside  the cell are deformed   and the constituent permittivity tensors are now anisotropic following \eqref{eq:photoelgen}. For example, in the background material we have
\begin{equation}
\label{eq:strainpermmatr}
\boldsymbol{\varepsilon}^\mathrm{str,m}_\mathrm{r}(\X) =\left[ \left(\boldsymbol{\varepsilon}_\mathrm{r}^\mathrm{unstr,m} \right)^{-1} + (\mathbf{p}^\mathrm{m} : \mathbf{s}(\X) )^{-1} \right]^{-1}.
\end{equation}

Third,   we return to the procedure outlined in Section \ref{sec:effn} and compute  the effective permittivity   for the strained cell  $(\varepsilon_{ij}^{-1})^\mathrm{str,eff}$ with strained constituent permittivities  (Figure \ref{fig:3figmerge}(c)).  A cubic crystal under an $s_{\mathrm{yy}}$ strain possesses   tetragonal symmetry, and so  the strained permittivity tensor is   uniaxial. Subsequently, two subwavelength Bloch vectors are necessary to calculate the effective tensor \eqref{eq:epseffsystem}. Comparing the change in inverse permittivity tensors for the imposed strain gives
\begin{equation}
p_\mathrm{xxyy}^\mathrm{eff} = -\frac{1}{D} \left[ \left( \boldsymbol{\varepsilon}^\mathrm{str,eff}_\mathrm{xx} \right)^{-1} -  \left( \boldsymbol{\varepsilon}^\mathrm{unstr,eff}_\mathrm{xx} \right)^{-1} \right].
\end{equation} 
 
\noindent We remark that for cubic crystals there are only three free parameters in the full photoelastic tensor, and the structure of the tensor means that only one simple strain need  be considered   to recover the effective photoelastic term for the SBS gain  \eqref{eq:gain}.

In  the event  that all  36 elements of the photoelastic tensor are required,   further imposed displacement fields must be considered.   For example,  the $p_\mathrm{yzyz}$ component can be obtained by applying  an $s_\mathrm{xz}$ strain  which corresponds to the   boundary conditions  
\begin{equation*}
\mathbf{u}\big|_{\pm \partial W_\mathrm{x,z}} = (D\mathrm{z} ,0,D \mathrm{x})^\mathrm{T}, \quad \mbox{and} \quad \mathbf{u} \cdot \mathbf{n} \big|_{\partial W_y} = 0,
\end{equation*}
however care must be taken to correctly determine the principal axes of the sheared cell.

\section{Numerical results} \label{sec:numer}
To illustrate the procedure outlined in previous sections, we   consider   simple cubic (sc) and face-centred cubic (fcc)  lattices of spheres in a background material, with one sphere per lattice site. We emphasise that these configurations are chosen as they admit closed-form expressions in the dilute limit, which is useful for independent validation \cite{smith2015electrostriction}. However, our method applies generally to all periodic structures for which an effective $\varepsilon_{ij}$ and $C_{ijkl}$ can be obtained.  Due to complexities  in numerical implementation (see Appendix \ref{sec:numimp}) we restrict our attention to filling fractions $f$ below the dense packing limit; explicitly, we consider the range  $0 <f<50\%$ for sc lattices and  $0 <f<73\%$ for fcc lattices. The dense packing  limit is $f = \pi/6$ for an sc lattice    and $f = \pi/(3\sqrt{2})$ for an  fcc lattice, below which the filling fractions are given explicitly by
\begin{equation}
f_\mathrm{sc} = \frac{4 \pi a^3}{3d^3}, \quad f_\mathrm{fcc} =\frac{16 \pi a^3}{3d^3},
\label{eq:fillfracs}
\end{equation}   
 where $a$ denotes the radius of the spherical inclusions and $d$ the period of the cubic lattice.    Above the dense packing limit, the filling fraction can be easily evaluated numerically.   In     the figures that follow,  the   SBS   gain  coefficient is   obtained by specifying  $\Omega = \Omega_\rmB$ in \eqref{eq:gain} to give
 \begin{equation}
 \label{eq:maxgain}
 \mathrm{max}\left( g_\mathrm{P} \right) = \frac{ 4 \pi^2  \gamma^2   }{   c_0 \lambda_1^2 n \rho   \nu_\mathrm{A} \Gamma^\mathrm{ }_\rmB}  ,
 \end{equation}
 where $\gamma = \varepsilon_r^2 p_\mathrm{xxyy}$ denotes the electrostrictive stress. We also examine   the  contribution from each term in \eqref{eq:maxgain} as the filling fraction is changed by considering:  
 \begin{widetext}
\begin{equation}
\label{eq:tenlogten}
10 \log_{10} \left( \frac{\mathrm{max}(\mathrm{g}_P)}{\mathrm{max}(\mathrm{g}_P^\rmb)} \right) = 
\underbrace{10 \log_{10} \left(  \left(\frac{\gamma}{\gamma^\rmb} \right)^2 \right) }_{\mbox{\footnotesize   electrostriction}}+ 
\underbrace{10 \log_{10}   \left(\frac{n^\rmb}{n}  \right) }_{\mbox{\footnotesize   refractive index}}+  
 \underbrace{ 10 \log_{10}   \left(\frac{\rho^\rmb}{\rho}  \right) }_{\mbox{\footnotesize   density}}+ 
\underbrace{10 \log_{10}   \left(\frac{\Gamma_\rmB^\rmb}{\Gamma_\rmB}  \right) }_{\mbox{\footnotesize   Brillouin linewidth}}+ 
\underbrace{10 \log_{10}   \left(\frac{\nu_\mathrm{A}^\rmb}{\nu_\mathrm{A}}  \right)}_{\mbox{\footnotesize   acoustic velocity}},
\end{equation}   
\end{widetext} 
\noindent where the  superscript $\rmb$ denotes the   bulk  parameter   value for the background material. In this way, the contribution from each term to the enhancement, or suppression, of the SBS gain coefficient can be straightforwardly determined by simply adding the value given by each curve at a specified filling fraction.

\begin{figure*}[t]
\centering
\includegraphics[width=0.287\linewidth]{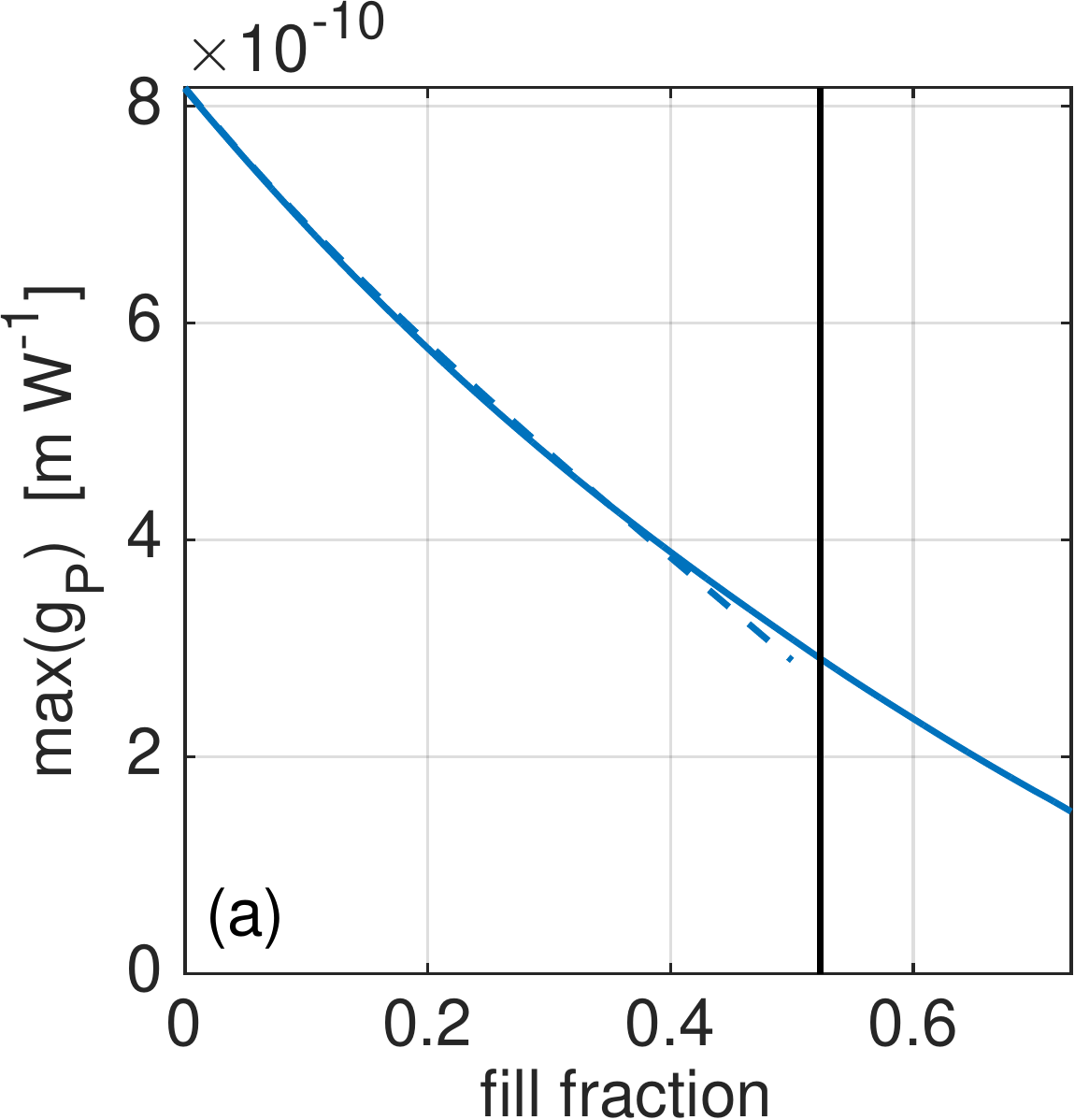} 
\includegraphics[width=0.369\linewidth]{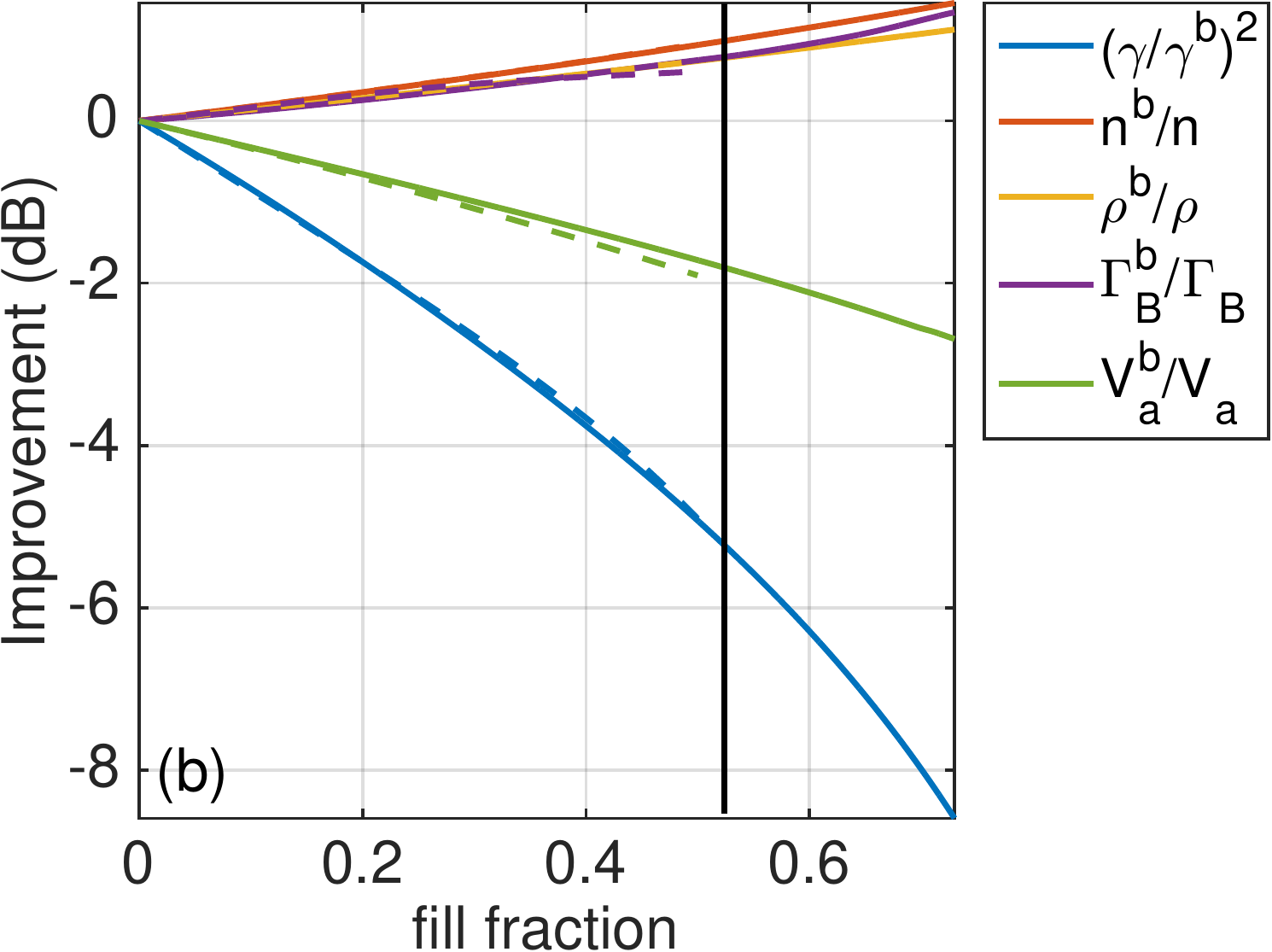}
\includegraphics[width=0.29\linewidth]{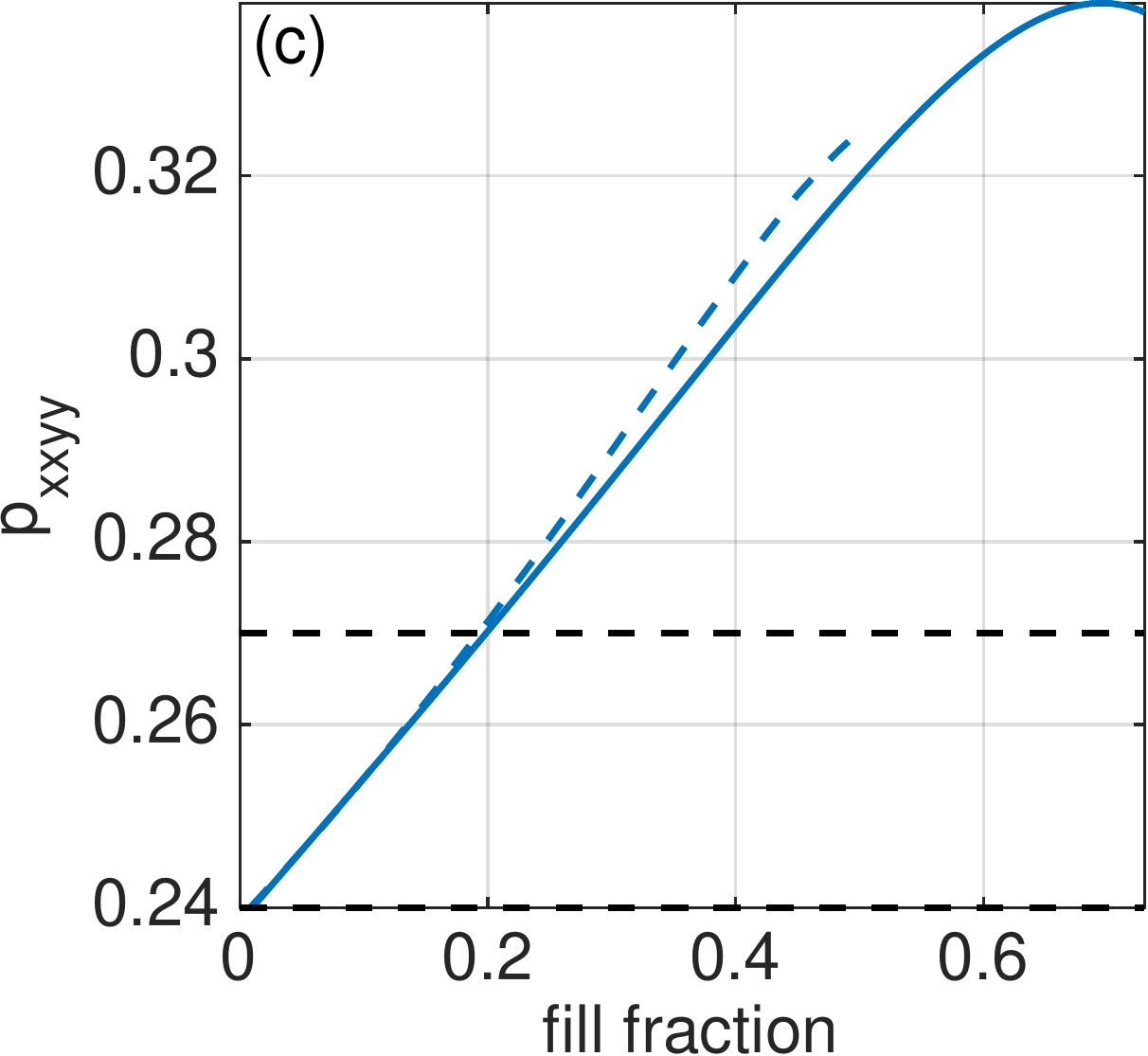}

\caption{    (a) Gain coefficient for  fcc lattice (solid) and sc lattice (dashed) of SiO$_2$ spheres in As$_2$S$_3$ at $\lambda_1 = 1550$ nm for        $d = 100$ nm (b)  contribution from each term in \eqref{eq:gain} to   improvement  in $g_\mathrm{P}$     (c) Photoelastic term $p_\mathrm{xxyy}$ for fcc lattice (blue) and sc lattice (dashed).  Solid black vertical lines denote   dense packing limit and horizontal black dashed lines denote intrinsic $p_\mathrm{xxyy}$ values. For   sc lattice at $f=50\%$ we find $\mathrm{max}(g_P) = 2.9 \times 10^{-10} \, \mathrm{m}\cdot\mathrm{W}^{-1}$, $\Omega_\rmB/(2\pi) = 9 \, \mathrm{GHz}$, $\Gamma_\rmB/(2\pi) = 28.7 \, \mathrm{MHz}$, $p_\mathrm{xxyy} = 0.32$, $n = 1.91$, $\rho = 2700 \, \mathrm{kg} \cdot \mathrm{m}^{-3}$, $V_\mathrm{A} = 3749 \, \mathrm{m} \cdot \mathrm{s}^{-1}$ and for   fcc lattice at $f=73\%$ we find $\mathrm{max}(g_P) = 1.5 \times 10^{-10} \, \mathrm{m}\cdot\mathrm{W}^{-1}$, $\Omega_\rmB/(2\pi) = 9.8 \, \mathrm{GHz}$, $\Gamma_\rmB/(2\pi) = 24.3 \, \mathrm{MHz}$, $p_\mathrm{xxyy} = 0.34$, $n = 1.69$, $\rho = 2470 \, \mathrm{kg} \cdot \mathrm{m}^{-3}$, $V_\mathrm{A} = 4489 \, \mathrm{m} \cdot \mathrm{s}^{-1}$. }
\label{fig:1}
\end{figure*}

 In Figure \ref{fig:1}a,  we show  the  gain coefficient as a function of filling fraction, for an sc and fcc lattice of \siot  spheres in an otherwise uniform  \astst background.  In this figure, we observe that the suppression of the  gain coefficient is   almost entirely   independent of the lattice configuration for $0<f<50\%$. This suggests that the result is independent of any lattice configuration (including random), since the metamaterial structuring is  both optically and acoustically subwavelength. At $f=50\%$ we achieve a $65\%$ suppression of the background gain coefficient  and at $f=73\%$ (fcc) we achieve a suppression of 82\%. As a result, the power threshold for SBS in \astst has been   raised considerably. Note that when calculating these percentages we have used    $\mathrm{max}(g_P) = 8.2 \times 10^{-10} \, \mathrm{m}\cdot\mathrm{W}^{-1}$  at $f=0\%$, which is within 10\% of experimentally obtained values for the gain coefficient in  \astst  waveguides \cite{pant2011chip}. A solid black vertical line is shown at $f=\pi/6$ which marks the dense packing threshold for sc lattices of spheres.

 In Figure \ref{fig:1}b we determine the contribution from each term present in \eqref{eq:maxgain}, as seen in \eqref{eq:tenlogten}, and observe that reduced electrostriction is the primary mechanism behind the suppressed gain coefficient, followed by the increased longitudinal acoustic velocity. All other terms   work to enhance the gain coefficient, however the electrostriction and acoustic velocity terms outstrip these contributions to suppress the gain coefficient overall. The solid curves denote   results for the fcc lattice and broken lines denote an sc lattice, where as in Figure \ref{fig:1}a, the difference is negligible between the two lattice types over $0<f<50\%$.

Although Figure \ref{fig:1}b demonstrates suppression in the electrostriction parameter for \siot spheres in \aststnb, the photoelastic constant $p_\mathrm{xxyy}$ is actually increasing, as can be seen in Figure \ref{fig:1}c. Here, we superimpose the result for sc (dashed blue curve) and fcc (solid blue curve) lattices, where dashed horizontal black lines denote the intrinsic  $p_\mathrm{xxyy}$ for the constituent materials. It is immediately apparent  that the photoelastic constant for the metamaterial is not given by a simple mixing of the two constituent values, but instead, also includes a strong   artificial photoelasticity contribution \cite{smith2015electrostriction} that arises due to the different mechanical responses of the   constituent media. As in the preceding figures, the difference between the sc and fcc lattice configurations is minor, and the   value of $p_\mathrm{xxyy} = 0.34$ at $f=73\%$ is extremely close to the maximum value obtained for $p_\mathrm{xxyy}$ at $f = 69\%$. The explanation for   the suppressed   electrostriction parameter lies in the reduced effective index of the two materials; although the effective photoelastic term is increasing strongly, the effective permittivity is decreasing at a faster rate, and so suppressed electrostriction results overall.

\begin{figure*}[t]
\centering
\includegraphics[width=0.305\linewidth]{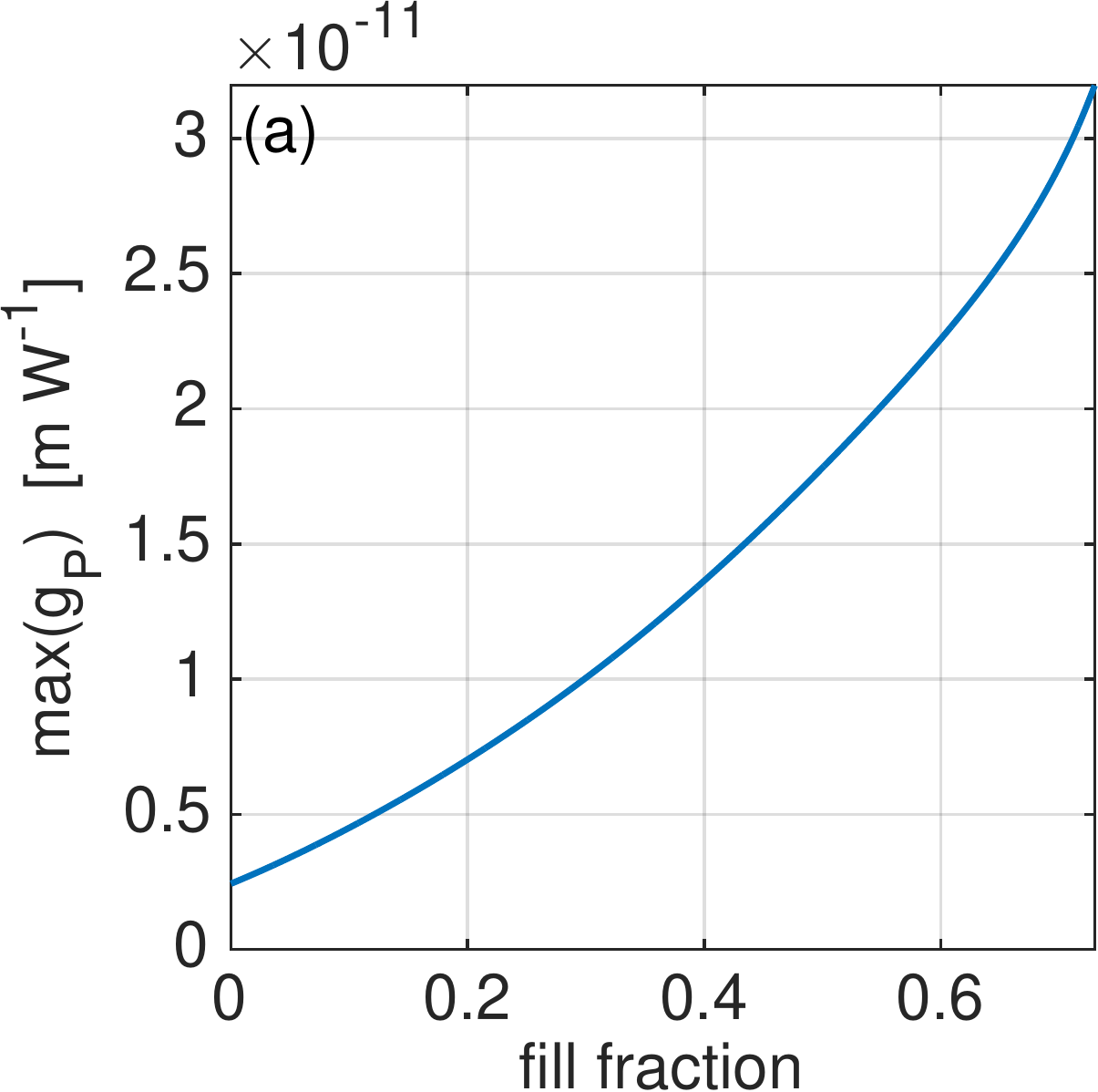} 
\includegraphics[width=0.369\linewidth]{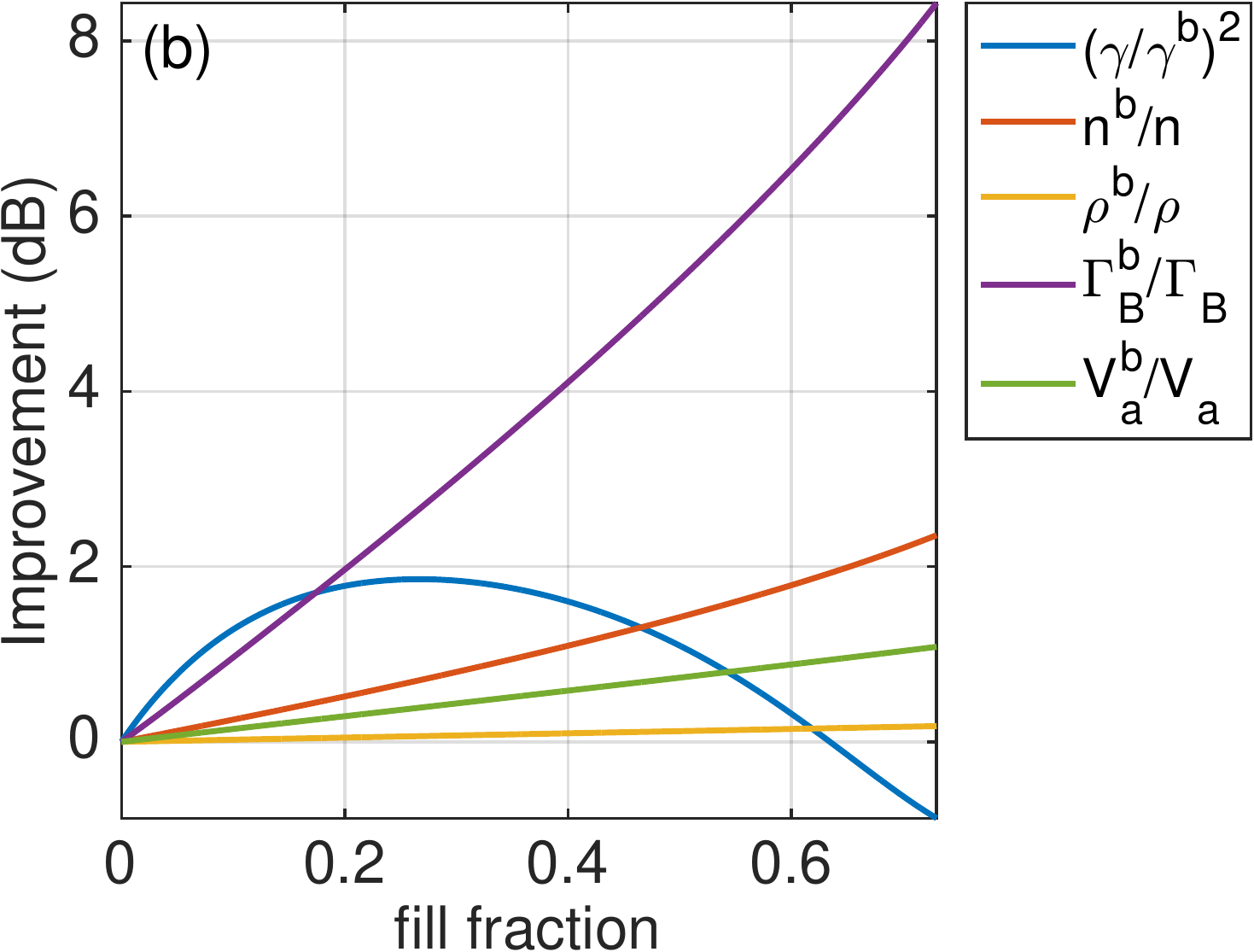}

\caption{(a) Gain coefficient for  fcc cubic lattice of SiO$_2$ spheres in Si  at $\lambda_1 = 1550$ nm for    $d = 100$ nm, (b) contribution from each term in \eqref{eq:gain} to   improvement  in $g_\mathrm{P}$     for SiO$_2$ spheres in Si. At $f = 73\%$ we find $\mathrm{max}(g_P) = 3.2 \times 10^{-11} \, \mathrm{m}\cdot\mathrm{W}^{-1}$ $\Omega_\rmB/(2\pi) = 17.1 \, \mathrm{GHz}$, $\Gamma_\rmB/(2\pi) = 45.9 \, \mathrm{MHz}$, $p_\mathrm{xxyy} = 0.13$, $n = 2.02$, $\rho = 2235 \, \mathrm{kg} \cdot \mathrm{m}^{-3}$, $V_\mathrm{A} = 6569 \, \mathrm{m} \cdot \mathrm{s}^{-1}$. Here $\mathrm{max}(\gamma) = 3.06$ at $f=27\%$}
 \label{fig:2}
\end{figure*}

In Figure \ref{fig:2}a, we show the maximum gain coefficient   for an fcc   lattice of \siot spheres in Si, which shows a monotonically increasing SBS gain coefficient from $\mathrm{max}(g_P) = 2.4 \times 10^{-12} \, \mathrm{m}\cdot\mathrm{W}^{-1}$ at $f=0\%$ to $\mathrm{max}(g_P) = 3.2 \times 10^{-11} \, \mathrm{m}\cdot\mathrm{W}^{-1}$ at $f=73\%$, and corresponds to more than one order of magnitude enhancement (an enhancement  of 13.3 compared to the background value). Referring to Figure \ref{fig:2}b,  it is immediately apparent that electrostriction is not the primary mechanism for the enhancement in the gain coefficient, it is instead the   Brillouin linewidth. Given that our earlier work \cite{smith2016control} suggests electrostriction is the {\it force majeure} behind the enhancement and suppression of the gain coefficient, this is an important demonstration that    all terms in \eqref{eq:maxgain} must be calculated  in order to determine the SBS properties of a composite material, and that it is incorrect in general to rely on  the electrostriction alone. In this example, the electrostriction achieves a maximum of $\gamma = 3.06$ at $f = 26\%$, returning to the  electrostriction value for the background material at $f=63\%$. Over the range $0<f<63\%$ all terms contribute positively to enhance the gain coefficient, after which the electrostriction acts to suppress the gain coefficient.

\begin{figure*}[t]
\centering
\includegraphics[width=0.285\linewidth]{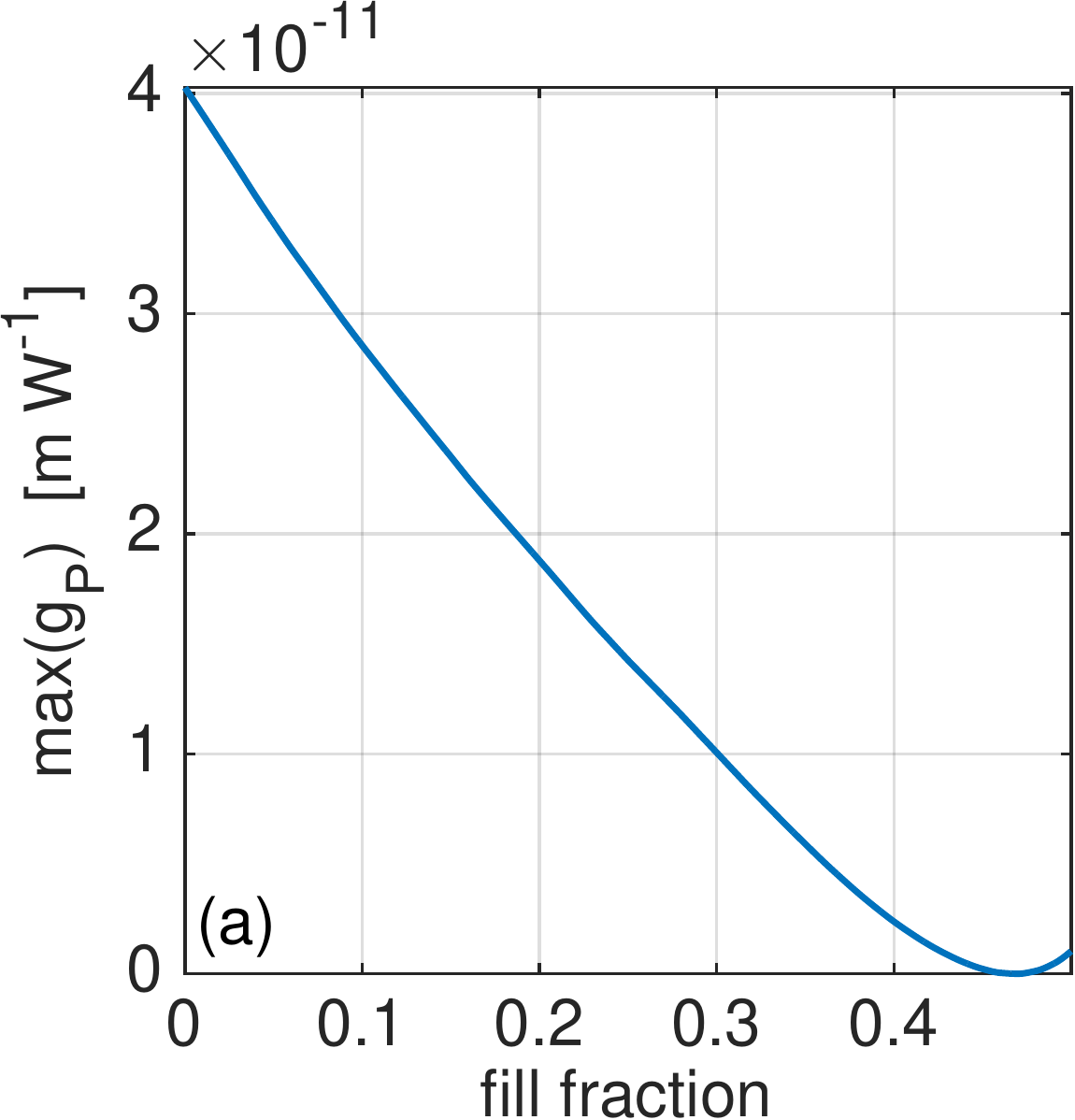} 
 \includegraphics[width=0.369\linewidth]{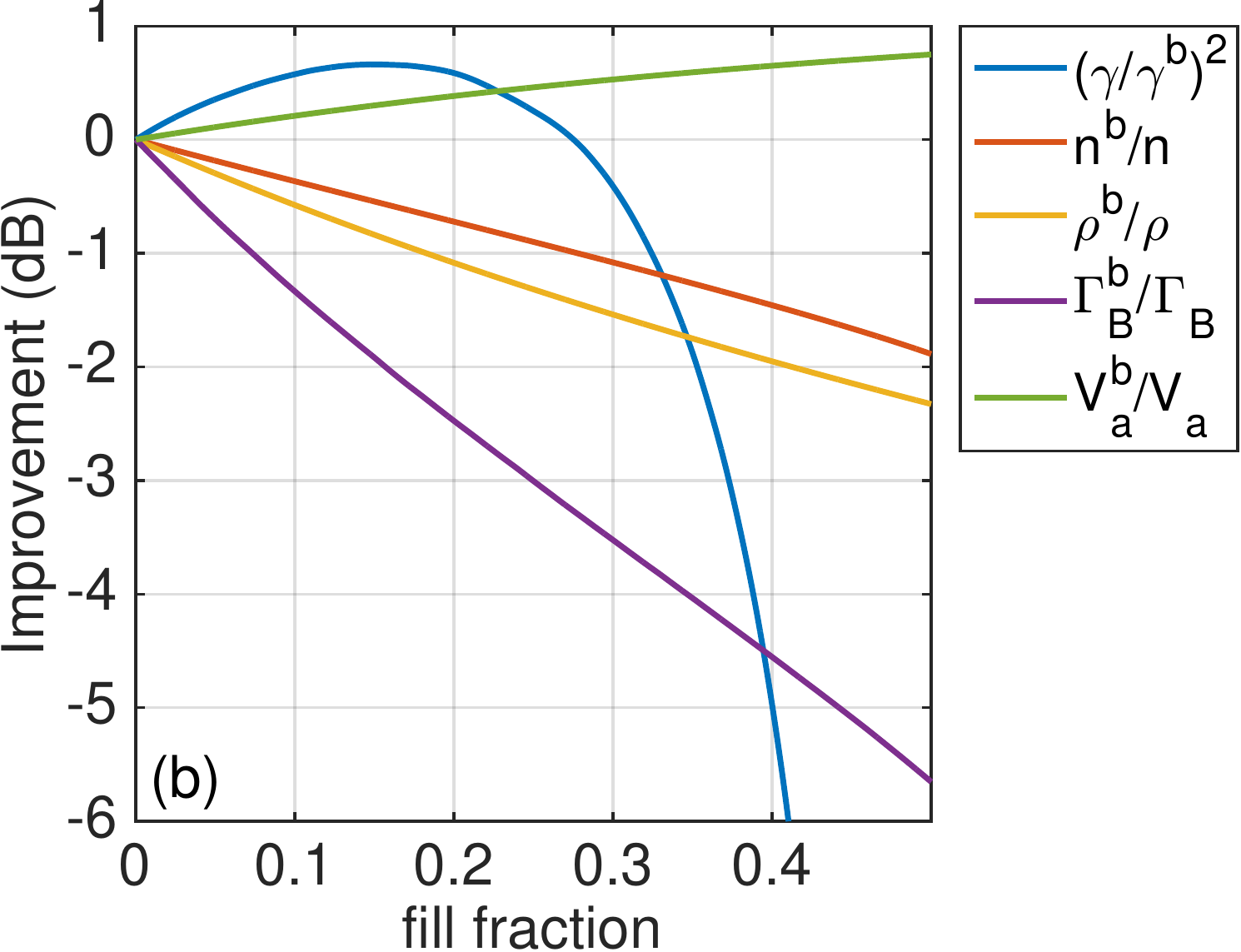}

\caption{(a) Gain coefficient for   sc lattice of GaAs spheres in SiO$_2$    at $\lambda_1 = 1550$ nm for    $d = 50$ nm, (b) contribution from each term in \eqref{eq:gain} to   improvement  in $g_\mathrm{P}$     for GaAs spheres in SiO$_2$. Here, $\mathrm{max}(g_P) = 0$ at $f=47\%$ and $\mathrm{max}(\gamma) = 1.28$ at $f=15\%$.}
\label{fig:3}
\end{figure*}

The final example  we consider is an  sc lattice of GaAs spheres in \siot as shown in Figure \ref{fig:3}. Here, the gain coefficient is completely suppressed at a filling fraction of $f=47\%$ and can be attributed to a perfectly vanishing photoelastic parameter. In this example, the increased linewidth acts as the primary mechanism behind the suppression of SBS from $0<f<39\%$, after which the vanishing electrostriction term dominates. From Figure \ref{fig:3}b, we observe that the electrostriction reaches a maximum of $\gamma = 1.28$ at $f = 15\%$ before returning to the background material value at $f=28\%$ and ultimately reaching a zero near $f=50\%$. The only parameter that consistently works against the suppression of the SBS gain is the acoustic velocity. Note that  at  $f=0\%$ we have $\mathrm{max}(g_P) = 4 \times 10^{-11} \, \mathrm{m}\cdot\mathrm{W}^{-1}$ which differs by $13\%$ from experimental data for the gain coefficient in fused silica waveguides   \cite{abedin2005observation}.

\section{Discussion  } \label{sec:concl}
We have presented a rigorous procedure for determining all material and wave propagation parameters that feature in the SBS gain coefficient for a metamaterial. This involves intensive numerical calculations for the   optical properties,   mechanical properties, and   photoelastic properties present in \eqref{eq:gain}. The implementation of routines to evaluate these parameters is complicated, as discussed in Appendix \ref{sec:numimp}, but have been successfully used    to demonstrate both enhancement and suppression of the SBS gain coefficient in composite media. We have demonstrated  that for arrays of spheres, the SBS gain is    independent of the cubic lattice configuration, provided the metamaterial is optically and acoustically subwavelength. Also, we have shown that electrostriction does not always completely determine the behaviour of the SBS gain coefficient; the Brillouin linewidth is shown to play an important role  for certain material combinations.

The methods     will enable    researchers to characterise  the SBS properties of   exotic metamaterials, which may open promising new paths for opto-acoustics and SBS. Materials with enhanced or suppressed gain coefficients are of particular interest for on-chip applications.

\begin{figure*}[t]
\centering
 \includegraphics[width=0.33\linewidth]{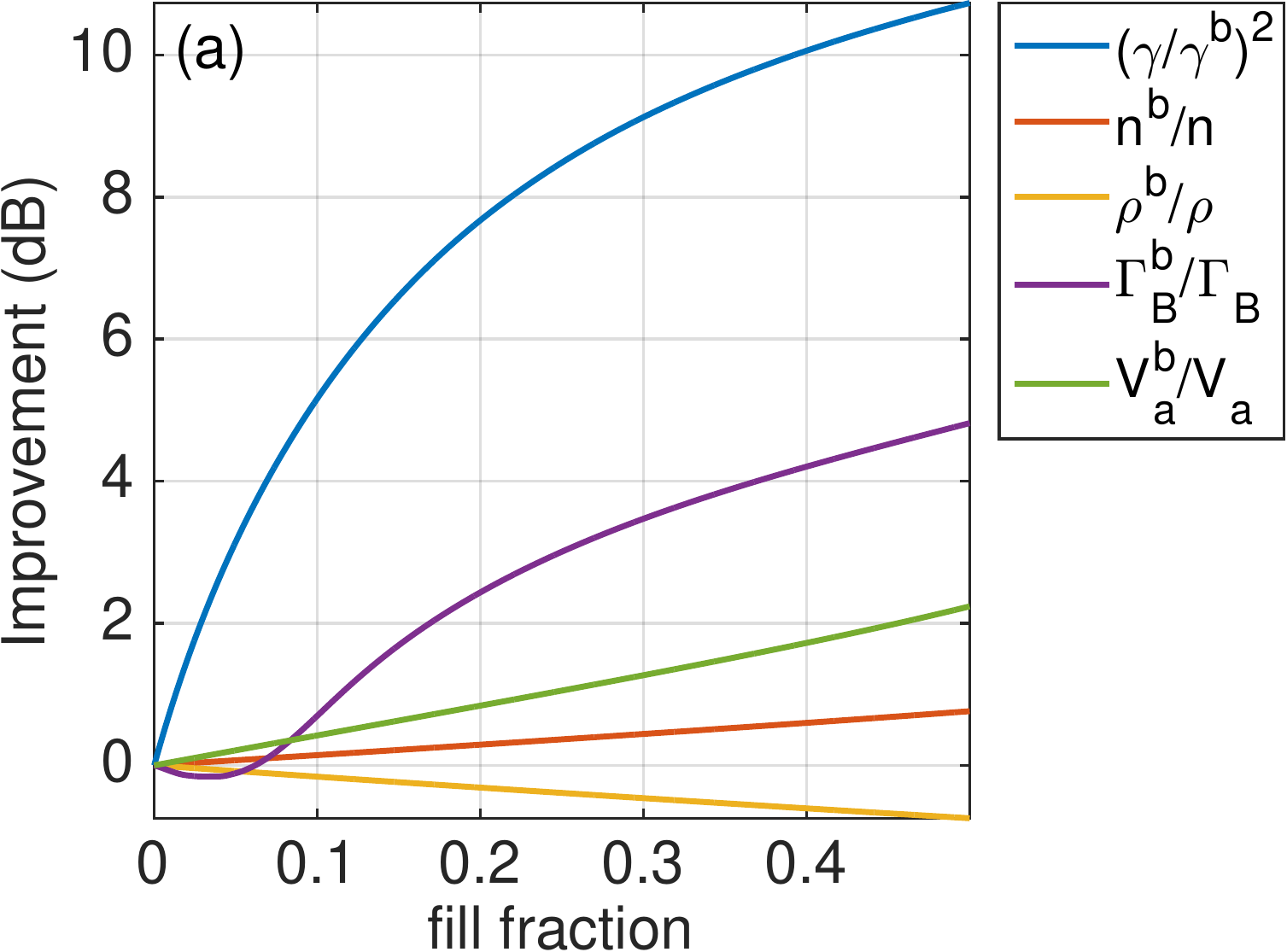} \;
 \includegraphics[width=0.27\linewidth]{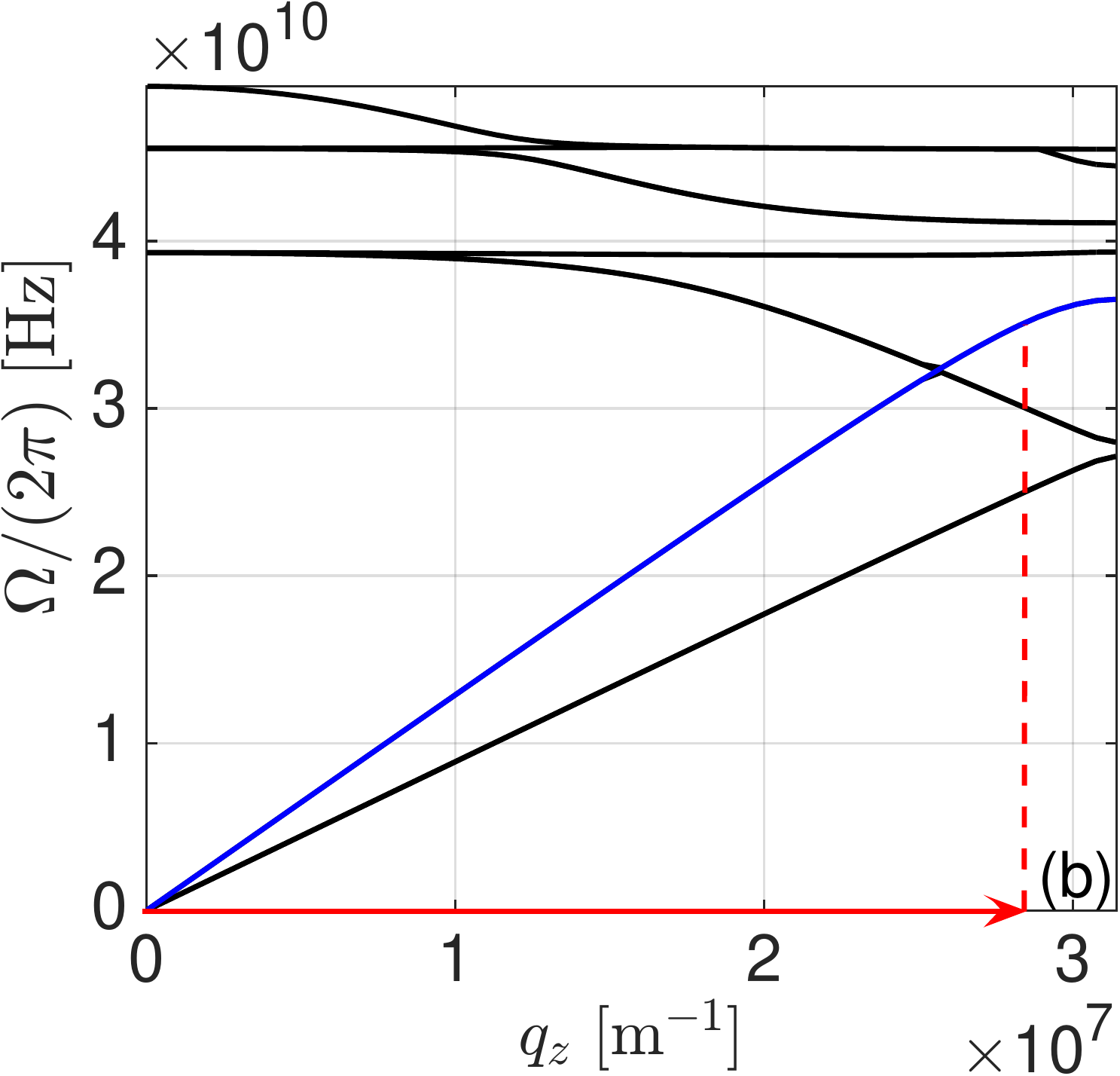} \;
 \includegraphics[width=0.27\linewidth]{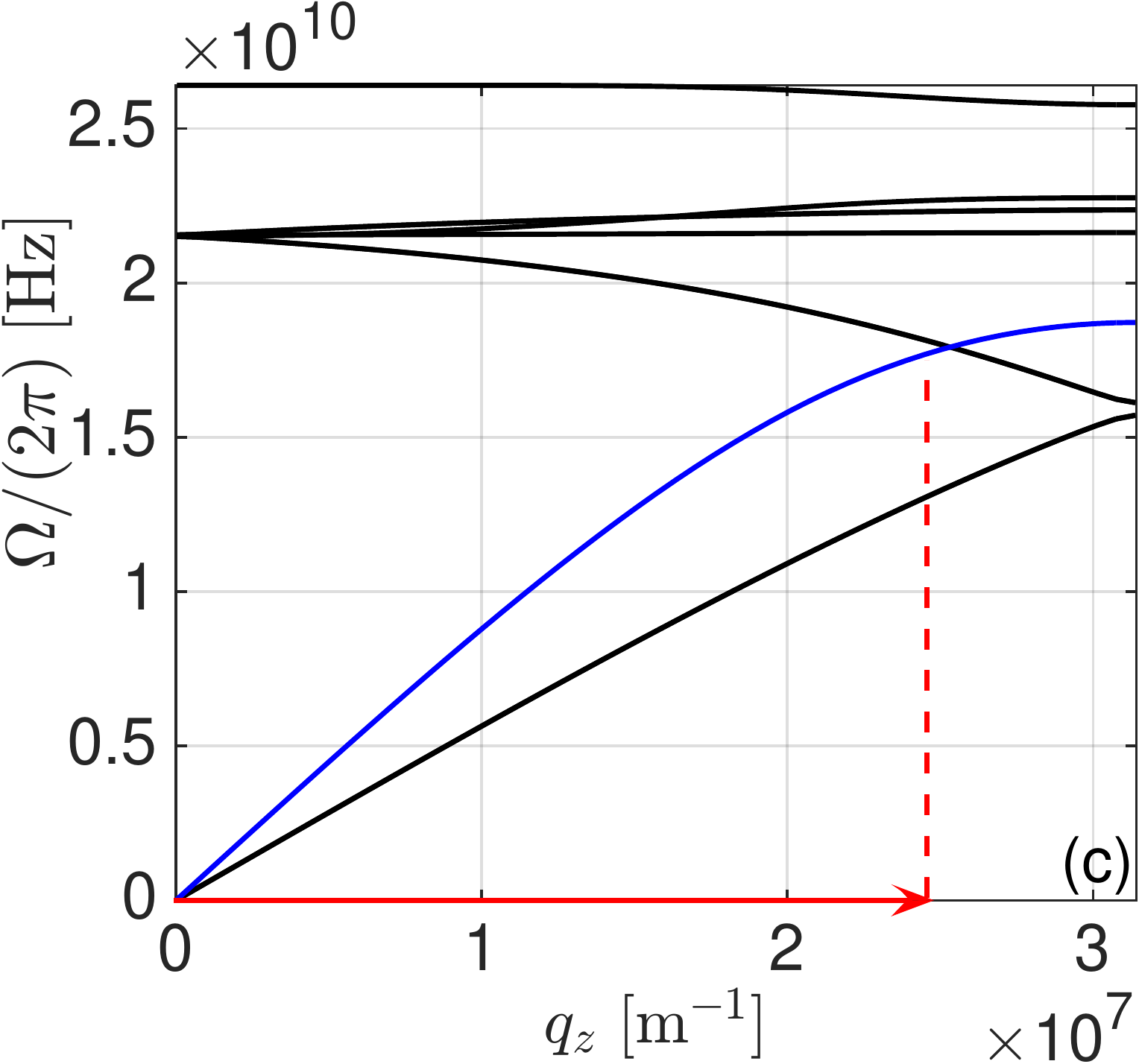}
 \includegraphics[width=0.33\linewidth]{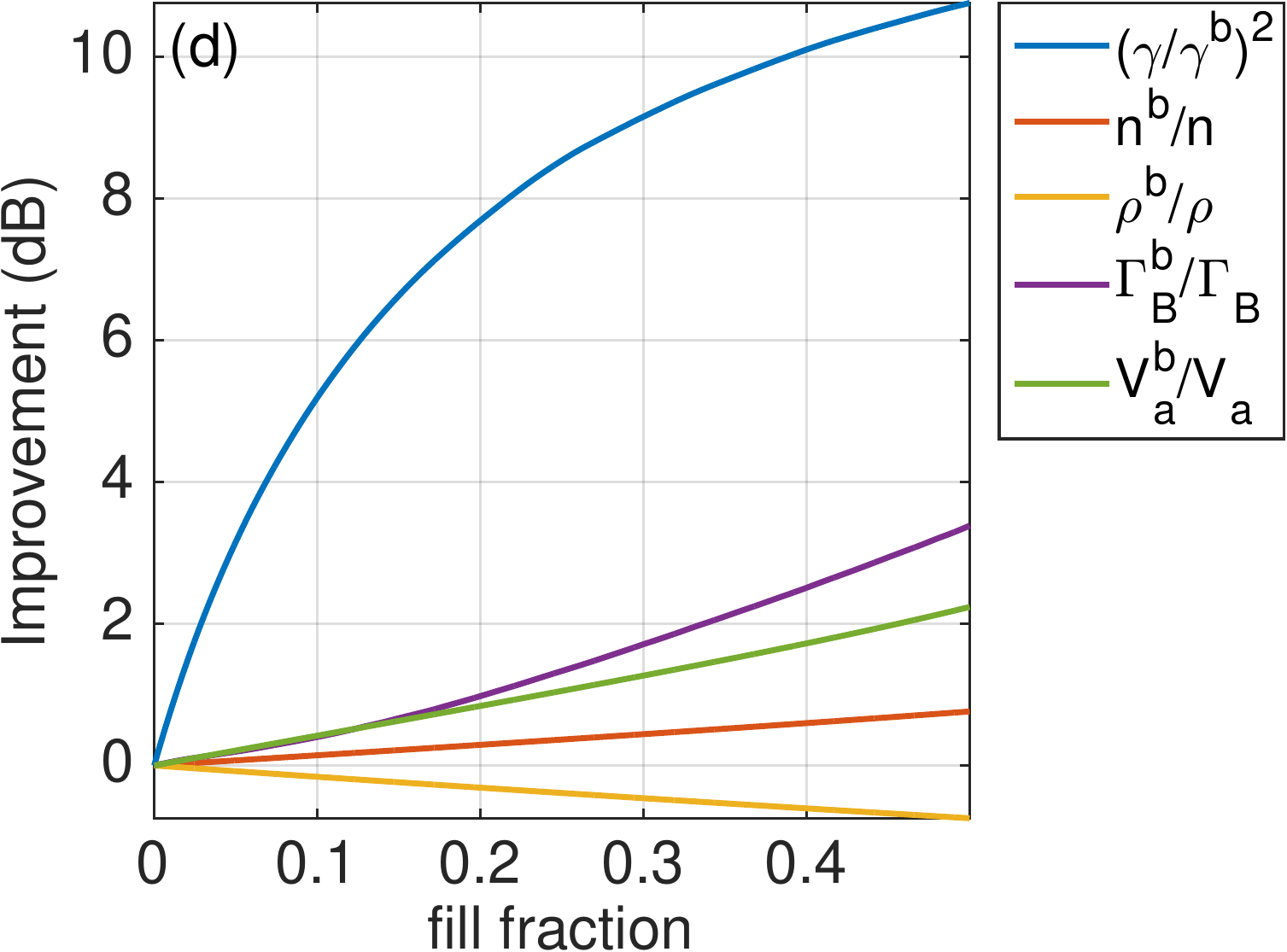} \;
 \includegraphics[width=0.27\linewidth]{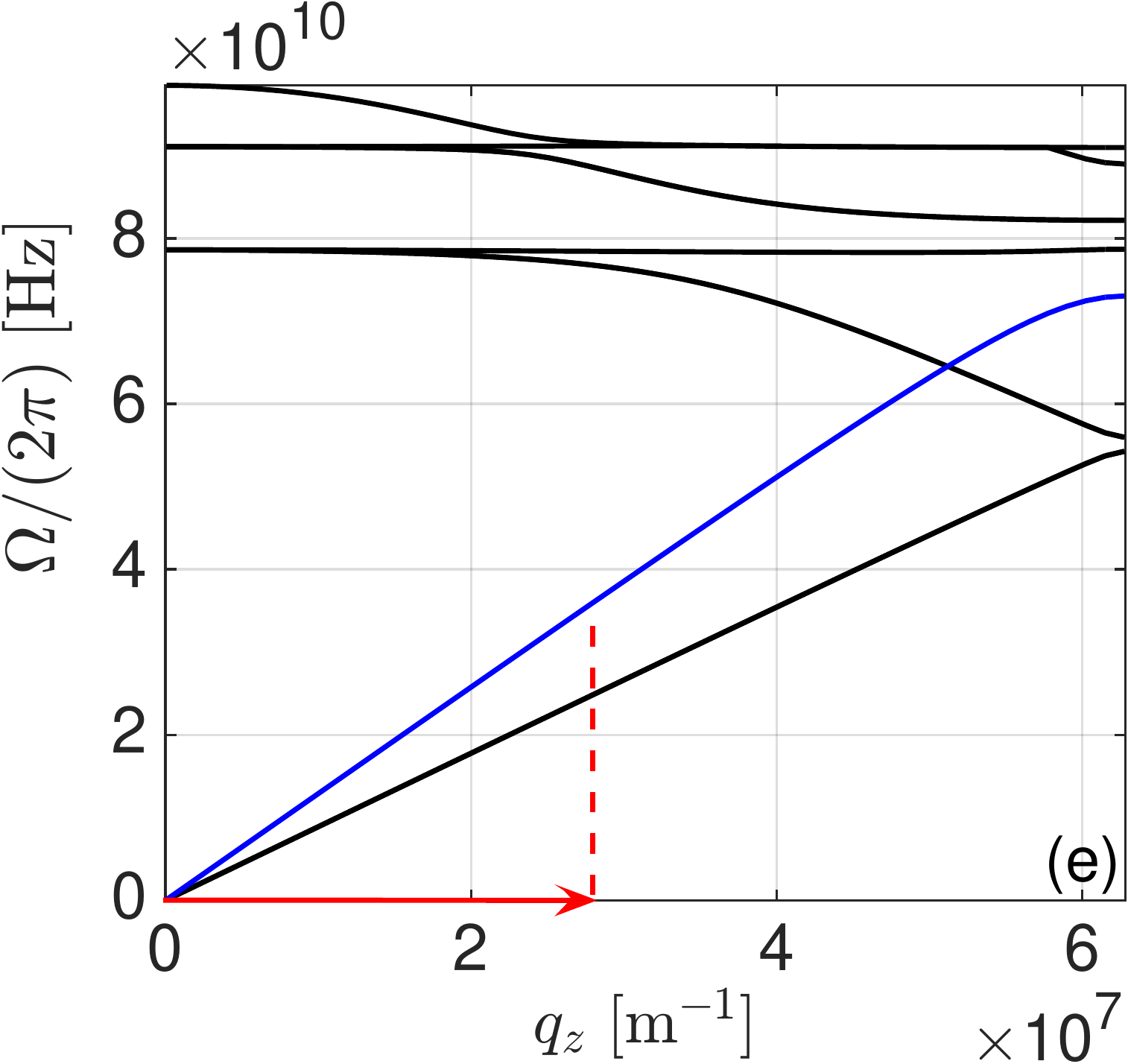} \;
 \includegraphics[width=0.27\linewidth]{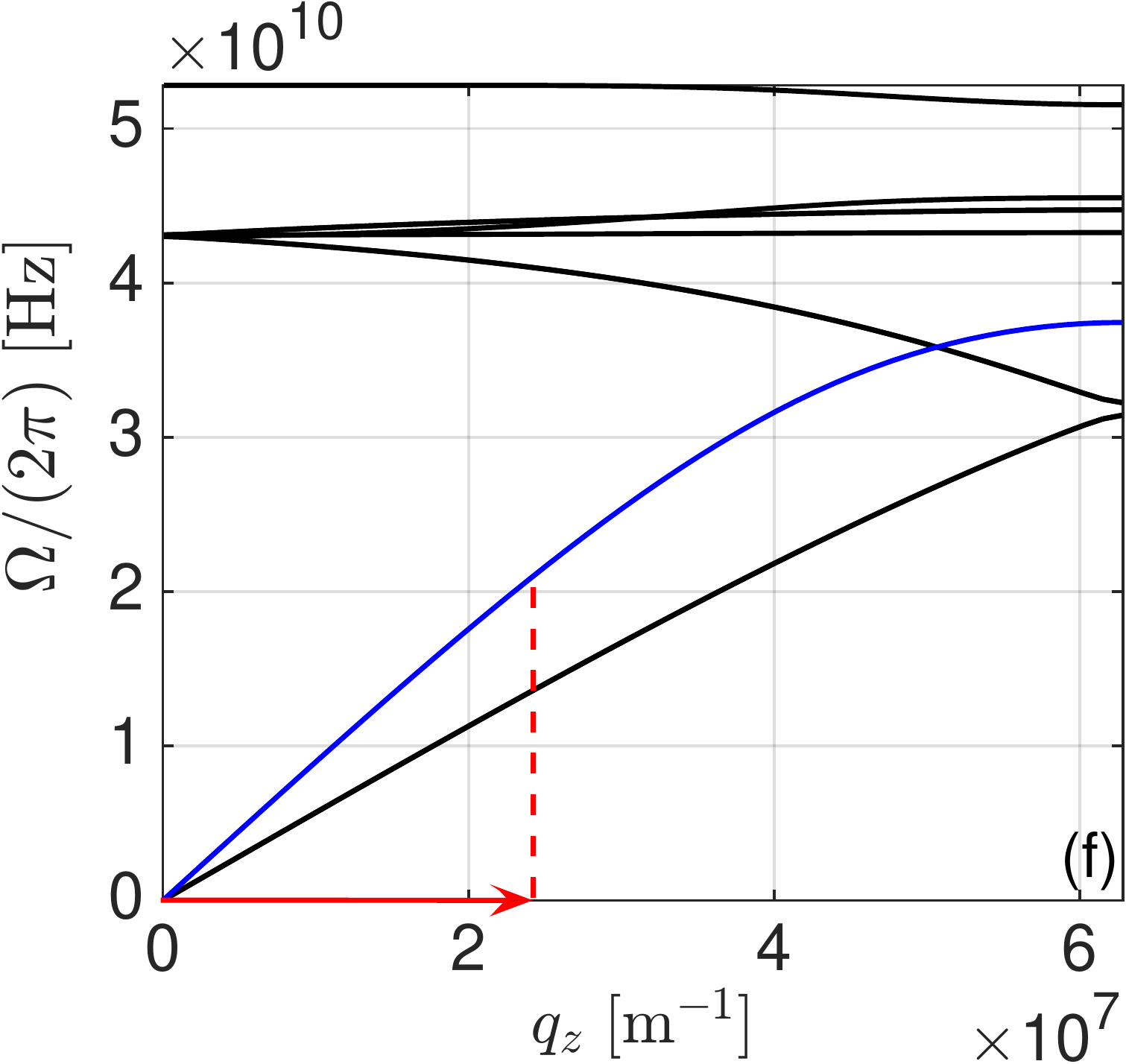}

 \caption{Satisfying   acoustic subwavelength condition for sc lattice of  \astst spheres in Si  at $\lambda_1 = 1550\, \mathrm{nm}$: (a) Erroneous results  for contribution from each term in \eqref{eq:gain} to   improvement  in $g_\mathrm{P}$ for $d=100 \, \mathrm{nm}$, (b) acoustic band structure   for  \astst spheres in Si at $f = 4\%$ for $d=100 \, \mathrm{nm}$ ($n_\mathrm{eff} = 3.4$) for $0<q_z<\pi/d$, with   SBS resonant wave vector $\mathbf{q}_\rmB$ (red arrow)   superposed and   first longitudinal band surface   highlighted (blue curve) where effective parameters can be obtained (c) acoustic band structure for \astst spheres in Si at $f=40\%$ ($n_\mathrm{eff} = 3.03$) for $d=100 \, \mathrm{nm}$ demonstrating   $\mathbf{q}_\rmB$ is no longer in linear dispersion regime.    Figures (d)--(f) are analogous to (a)--(c) but correspond to $d=50 \, \mathrm{nm}$ where (d) is reproduced from \cite{smith2016control} and $\mathbf{q}_\rmB$ now lies in   linear dispersion regime  about the centre of the first Brillouin zone.  }  
\label{fig:4}
\end{figure*}

We note that the quasi-static assumptions necessary for a number of steps in our procedure   require  that  we have a sufficiently large number of unit cells per   optical and acoustic wavelength in the material. It is only in this setting that the descriptions presented in this paper are valid. To demonstrate this point we include Figure \ref{fig:4} where results for an sc lattice of \astst spheres in Si for $d=100 \, \mathrm{nm}$ and $d=50 \, \mathrm{nm}$ can be found. In Figure \ref{fig:4}a we observe that the acoustic parameter $\Gamma_\rmB$ decreases    at first, to reach  a local minimum at $f=4\%$, before increasing in a predictable manner. The explanation for this behaviour can be found in Figures \ref{fig:4}b and \ref{fig:4}c where we compute the acoustic band structure at $f=4\%$ and $f=40\%$ filling fractions. The high effective refractive index for these structures means that the magnitude of the   SBS resonant wave vector $\mathbf{q}_\rmB$ is large. Consequently, the acoustic wave vector is no longer sufficiently close to the  $\Gamma$ point and so results are incorrect. Reducing the lattice period to $d = 50\, \mathrm{nm}$ for example, scales the band structure of the material such that the acoustic wave vector  is then within a neighbourhood of $\Gamma$ where effective acoustic parameters can be computed (Figures \ref{fig:4}d--\ref{fig:4}f). Note that if the period of the lattice $d$ is sufficiently large, then   $|\mathbf{q}_\rmB|>\pi/d$ and we exit the first Brillouin zone, potentially giving rise to acoustic Umklapp scattering.

We also remark that considerable care must be taken to ensure   the symmetry properties of   effective  material tensors are correctly characterised:  the symmetry of the underlying lattice (i.e., cubic), the symmetry of the inclusion geometry, or structure inside the cell, and the symmetry of the constituent  tensors  (i.e., the permittivity or stiffness tensor), are all relevant in determining  the symmetry of the effective tensor.

Although not directly required for evaluating the   SBS gain coefficient, the   viscosity tensor for a  metamaterial  can also be evaluated using    energy density methods. In order to determine this tensor  we use the dissipative power,   which represents the rate of energy loss in the material, in place of the acoustic  energy density \eqref{eq:acenden}. The dissipative power  is given by $\partial_t \mathcal{E}_a$, or $\rmi \Omega \mathcal{E}_a$, which   amounts to a careful scaling of the   approach used to determine $C_{ijkl}^\mathrm{eff}$.

\section*{Acknowledgements}
This work was supported by the Australian Research Council (CUDOS Centre of Excellence, CE110001018).

\appendix

\section{    Definition of the  permittivity tensor} \label{sec:effperm}
In this section we   discuss   the   relationship between the    methods outlined here for characterising the optical properties of the structured material; the energy density method for   the effective permittivity, and  using the   slope of the first band surface near the $\Gamma$ point to retrieve the effective refractive index.

We begin by considering the electromagnetic energy density differential \cite{landau1984electrodynamics}
\begin{equation}
\rmd \mathcal{E} = \mathbf{D} \cdot \rmd \mathbf{E} + \mathbf{B} \cdot \rmd \mathbf{H} 
\end{equation}
from which   the electric displacement is  given by
\begin{equation}
\label{eq:DiHf}
D_i = \frac{\partial \mathcal{E}}{\partial E_i} \bigg|_{\mathbf{H}},
\end{equation}
which assumes no perturbation to the $\mathbf{H}$ field (no change to the magnetic susceptibility). Using the definition of the energy density $\mathcal{E} = |\mathbf{P}|/v_g$ where $\mathbf{P}$  is the Poynting vector and $v_g$ the group velocity  \cite{auld1973acoustic}, it   follows from \eqref{eq:DiHf} that
\begin{equation}
\label{eq:Didef}
D_i = \frac{1}{\mu_0 v_g} \frac{\partial | \mathbf{E} \times \mathbf{B} |}{\partial E_i} \bigg|_{\mathbf{H}}= \frac{|\mathbf{e}_i \times \mathbf{E}|}{\mu_0 v_g},
\end{equation}
where $\mathbf{e}_i$ denotes a Cartesian basis vector.  Next we assume that the material exhibits a linear optical response
\begin{equation}
\varepsilon_0 \varepsilon_{ij} = \frac{D_i}{E_j},
\end{equation}
 where for simplicity, we   consider an isotropic and homogeneous material which admits
\begin{equation}
\label{eq:finaleii}
\varepsilon_{ii} = \frac{1}{\varepsilon_0}\frac{D_i}{E_i} =  \frac{1}{\varepsilon_0 \mu_0 \omega v_p}  \frac{|\rme_i \times (\mathbf{k} \times \mathbf{E}) |}{E_i} = \frac{|\mathbf{k}| c_0^2}{\omega v_g} = \frac{c_0^2}{v_g v_p},
\end{equation}
where $v_p$ is the phase velocity. Thus, the   permittivity of an isotropic material is given by the inverse product of the group and phase velocities of the medium. In the long wavelength limit $v_g =v_p$ and so it follows that $n_\mathrm{eff} = c_0|\mathbf{k}|/\omega$.   For  an anisotropic and homogeneous material the energy density method   works   when the unit vector $\mathbf{e}_i$ is oriented along one of the principal axes of the permittivity tensor. Along other directions this treatment   fails, which is obvious considering the complex geometries of the isofrequency contours   for optically biaxial media, for example.

\section{ Numerical implementation of the effective photoelasticity method} \label{sec:numimp}
All   results presented   in the present work were obtained using COMSOL $4.4$. We emphasise that COMSOL was not developed with this application in mind, and so we outline the procedure to obtain $p_\mathrm{xxyy}^\mathrm{eff}$ for a cubic metamaterial, following the schematic in Figure \ref{fig:3figmerge}.

Step (a): the procedure for calculating  the effective permittivity of an unstrained unit cell is   straightforward. Implementing an `Eigenfrequency Study' inside the `RF module'    for $|\mathbf{k}| \ll \pi/d$ gives a desired set of eigenfrequencies, and we compute \eqref{eq:unifeffperm} using   the electric field mode corresponding to the smallest genuine eigenvalue (note the TE and TM modes are degenerate along $k_z$ in the vicinity of the  $\Gamma$ point). There are two issues when using the standard solver; the first is that an estimate for the lowest eigenfrequency must be specified. Even after specifying this shift, the lowest eigenvalues are often spurious modes  that  must be discarded. The second issue is that the Bloch vector $\mathbf{k}$ cannot be arbitrarily small; we recommend specifying $\mathbf{k} = (0,0,\pi \epsilon / d)$ where $\epsilon$ is a small parameter, and  scaling the small parameter until $\mathbf{k}$ is sufficiently close to the $\Gamma$ point (for example, $\epsilon =0.001$). As mentioned in Section \ref{sec:effn}, the optical wave equation \eqref{eq:maxwell}  for the full cell imposes Bloch conditions on all boundaries of the unit cell $\partial W$, however the convergence of eigenfrequencies using the full cell is     slow and requires considerable computational power. Convergence     can be significantly accelerated by using a reduced unit cell geometry.

Specifically, if we consider metamaterials with cubic symmetry and principal axes parallel to   a Cartesian basis vector (i.e., $\mathbf{k} \propto \mathbf{e}_\mathrm{x}$)  then  we can exploit the  symmetry of the problem in order to reduce our computational domain and thereby accelerate convergence. In other words, we halve the geometry of the unit cell  in each of the remaining Cartesian basis vector directions (i.e., $\mathbf{e}_\mathrm{y}$ and $\mathbf{e}_\mathrm{z}$), leaving $1/4$ of the original cell.  Depending on whether the TE or TM polarised solutions are sought, we   impose PEC conditions on the edges of the reduced cell in one direction (i.e., on $ \pm\partial W^\prime_\mathrm{y}$ where $\partial W^\prime$ denotes the boundary of the reduced cell) and PMC conditions in the remaining direction (i.e., on $\pm \partial W^\prime_\mathrm{z}$), with Bloch conditions imposed on  $\pm \partial W_\mathrm{x}^\prime$. Since we are  sufficiently close to the $\Gamma$ point (i.e., $|\mathbf{k}| \ll \pi/d$),   the modes for both TE and TM polarisations are degenerate at long wavelengths. Accordingly, provided one reduced boundary has PMC conditions and the other has PEC, then the correct effective permittivity is obtained.

 Step (b): The strained reduced unit cell and corresponding internal strain field are obtained through a Stationary Study of a linear elastic material in the Solid Mechanics module  of COMSOL 4.4. We emphasize strongly that the calculation outlined here is one of a strain and not of a stress on a three-dimensional unit cell. Since the problem is symmetric for the simple strain we consider, the unit cell geometry can also be reduced, which formally restrict solutions to one symmetry class (i.e., longitudinal solutions). For the reduced unit cell, the boundary conditions in the reduced directions (i.e., on $\partial W^\prime_\mathrm{x}$ and $\partial W^\prime_\mathrm{z}$) are vanishing normal displacement $\mathbf{u} \cdot \mathbf{n} = 0$. To induce an $s_\mathrm{yy}$ strain, we impose the  displacements   \eqref{eq:syydispcond} on $\partial W_\mathrm{y}^\prime$. From the longitudinal mode $\mathbf{u}$  a strain tensor $s_{ij}$ can be computed inside the unit cell. This strain field is exported as several text files for each domain, along with the domain number. We then import these data files into MATLAB so that the strained permittivity tensor inside each deformed domain  can be calculated as in \eqref{eq:strainpermmatr}. The meshing for this problem must exploit all possible symmetries to ensure numerical stability in Step (c). Note that an artificial boundary must be introduced at $y=0$ and the   mesh for all boundaries with $\mathbf{n} = \mathbf{e}_y$ must be identical.

 Individual  text files corresponding to each domain are generated in MATLAB, which contain  the strained permittivity tensor as a function of discretized $(x,y,z)$ coordinates inside each domain. This is done to ensure  the permittivity tensor is uniquely and correctly assigned in Step (c) without encountering interpolation issues across domain boundaries. We remark that the strained unit cell geometry can only be exported when `Geometric Nonlinearity' is included in the model. With this functionality enabled, the strained unit cell is exported as a deformed mesh file using the `Remesh deformed configuration' option (as an mphtxt file). The geometric nonlinearity will calculate different stress and strain tensors  and so care must be taken to use the Engineering strain tensor for our analysis, as this corresponds to the linear theory treatment presented here. In practice,  there is little difference between the linear and nonlinear strain tensors for unit cells that are strongly subwavelength and subject to small strains.

Step (c): This   repeats the processes outlined in Step (a), except that we import the geometry (the mphtxt file) and define all six elements of the permittivity tensor $\varepsilon_{ij}$ in each domain using interpolated functions. Even if the strained unit cell is computed with an extremely dense mesh, it is still often necessary to `repair' the geometry, until COMSOL   recognises the correct number of boundaries and domains. This is a nontrivial step and the `repair tolerance' must be carefully chosen.  Once  all six permittivity tensor elements $\varepsilon_{ij}(x,y,z)$ are imported and defined inside each and every domain as interpolated functions, these  must then be  assigned to a `Material'   for each domain. Solving the optical wave equation \eqref{eq:maxwell} for the reduced and strained cell configuration,  we obtain the first   linear equation of the two equations necessary to determine the effective permittivity tensor (see \eqref{eq:epseffsystem}). To obtain the second linear equation using a reduced cell geometry, it is necessary to solve the optical wave equation  problem for $\mathbf{k}\propto \mathbf{e}_\mathrm{z}$. Accordingly, we use the Geometry tools in COMSOL to delete half of the existing cell geometry, copy the remaining geometry, and then reflect the copied geometry about $\mathbf{z}=0$ to obtain the appropriate reduced cell. Assigning  Materials classes to the reflected domains using $\varepsilon_{ij}(x,y,-z)$ where appropriate, and imposing PMC and PEC conditions on the   reduced directions ($\pm \partial W_\mathrm{x}^\prime$ and $\pm \partial W_\mathrm{y}^\prime$), we obtain the final eigenvalue problem. Solving this for   $|k_z| \ll 1 $ we evaluate  \eqref{eq:epseffsystem} once more to obtain the second linear equation, and subsequently the strained effective tensor.

\bibliography{SBS_meta_bib}

\end{document}